%% file: main.tex
\documentclass[conference]{IEEEtran}
\IEEEoverridecommandlockouts
\usepackage{cite}
\usepackage{amsmath,amssymb,amsfonts}
\usepackage{algorithmic}
\usepackage{graphicx}
\usepackage{textcomp}
\usepackage{xcolor}
\usepackage{url}
\usepackage{tikz}
\usepackage{listings}
\usepackage{authblk}
\usepackage{DejaVuSansMono}
\usepackage[T1]{fontenc}

\def\BibTeX{{\rm B\kern-.05em{\sc i\kern-.025em b}\kern-.08em
    T\kern-.1667em\lower.7ex\hbox{E}\kern-.125emX}}
    
\newcommand*\mycirc[1]{%
  \begin{tikzpicture}[baseline=(C.base)]
    \node[fill=black,text=white,draw,circle,inner sep=0.5pt](C) {#1};
  \end{tikzpicture}}
  
\begin{document}

\lstset{language=C,
basicstyle=\ttfamily\lst@ifdisplaystyle\fontsize{9}{11}\else\fontsize{9.5}{\baselineskip}\fi\selectfont,
commentstyle=\ttfamily\lst@ifdisplaystyle\fontsize{9}{11}\else\fontsize{9.5}{\baselineskip}\fi\fontshape{it}\selectfont,
keywordstyle=\ttfamily\lst@ifdisplaystyle\fontsize{9}{11}\else\fontsize{9.5}{\baselineskip}\fi\fontseries{b}\selectfont,
showstringspaces=false
}

\title{Analyzing Machine Learning Workloads Using a Detailed GPU Simulator}
\author[*]{\rm Jonathan Lew}
\author[*]{\rm Deval Shah}
\author[**]{\rm Suchita Pati}
\author[*]{\rm Shaylin Cattell}
\author[$\dag$]{\rm Mengchi Zhang}
\author[*]{\rm Amruth Sandhupatla}
\author[*]{\rm \\Christopher Ng}
\author[*]{\rm Negar Goli}
\author[**]{\rm Matthew D. Sinclair}
\author[$\dag$]{\rm Timothy G. Rogers}
\author[*]{\rm Tor M. Aamodt}
\affil[*]{Electrical and Computer Engineering, University of British Columbia}
\affil[ ]{\{jonathan.lew, shaylin.cattell, cng123\}@alumni.ubc.ca}
\affil[ ]{\{devalshah, amruth, negargoli93, aamodt\}@ece.ubc.ca}
\affil[**]{Computer Science, University of Wisconsin-Madison}
\affil[ ]{\{spati, sinclair\}@cs.wisc.edu}
\affil[$\dag$]{Electrical and Computer Engineering, Purdue University}
\affil[ ]{\{zhan2308, timrogers\}@purdue.edu}
\maketitle

\input{abstract}


\input{intro}
\input{background}
\input{implementation} 
\input{correlation}
\input{caseStudy}
\input{future}
\input{conc}

\bibliographystyle{IEEEtran}
\bibliography{ref}


\end{document}

%% file: abstract.tex
\begin{abstract}
Most deep neural networks deployed today are trained using GPUs via high-level frameworks such as TensorFlow \cite{tf} and PyTorch~\cite{paszke2017automatic}. This paper describes changes we made to the GPGPU-Sim simulator~\cite{aamodt2012gpgpu,bakhoda2009analyzing} to enable it to run PyTorch by running PTX kernels included in NVIDIA's cuDNN~\cite{chetlur2014cudnn} library.  We use the resulting modified simulator, which has been made available publicly with this paper\footnote{Source code available at \href{https://github.com/gpgpu-sim/gpgpu-sim\_distribution/tree/dev}{https://github.com/gpgpu-sim/gpgpu-sim\_distribution/ (dev branch)}}, to study some simple deep learning workloads.  
With our changes to GPGPU-Sim's functional simulation model we find GPGPU-Sim performance model running a cuDNN enabled implementation of LeNet for MNIST reports results within 30\% of real hardware. Using GPGPU-Sim's AerialVision performance analysis tool we observe that cuDNN API calls contain many varying phases and appear to include potentially inefficient microarchitecture behavior such as DRAM partition bank camping, at least when executed on GPGPU-Sim's current performance model.  
\end{abstract}

%% file: intro.tex
\section{Introduction}
\label{sec:intro}
\vspace{-1ex}

Machine learning is being employed to tackle a rapidly growing set of problems.  In recent years deep neural networks (DNNs) have made striking advances in accuracy.  Training DNNs requires massive amounts of computational power, which is currently predominantly done with graphics processor units (GPUs).  While industry has rapidly introduced changes to GPU architectures to support machine learning training, such as Tensor Cores and NVLINK introduced in the NVIDIA Volta architecture~\cite{nvidia_volta}, academic researchers have largely focused on designing inference accelerators.  Although the focus of academic researchers is to exploit
 the strong potential for neural network deployment in mobile platforms (e.g., iPhone X, Huawei) and small embedded devices~\cite{howard2017mobilenets, iandola2016squeezenet,wu2016quantized,rastegari2016xnor}, another reason for the lack of academic research on optimizing GPUs for machine learning may be the lack of support in current architecture simulators for running these workloads.  This paper takes an important step towards addressing this shortcoming.

Popular machine learning frameworks such as TensorFlow and PyTorch typically expose a high-level python application programming interface (API) to developers.  Calls to this API invoke computation on a GPU via specialized precompiled libraries such as cuBLAS~\cite{nvidia2008cublas} and cuDNN~\cite{chetlur2014cudnn}.  To achieve the highest levels of performance these libraries are typically provided by hardware vendors.  These libraries take advantage of the vendor's detailed knowledge of their product's microarchitecture, which is typically not fully described in publicly available documentation.  As a result, popular open source GPU architecture simulators such as GPGPU-Sim~\cite{bakhoda2009analyzing,aamodt2012gpgpu} are unable to run applications that make use of these precompiled libraries.  Indeed, we confirmed with the maintainers of GPGPU-Sim that a key limitation of the currently available version of GPGPU-Sim is the lack of support for applications that use precompiled libraries.  
In this paper, we focus on enabling support for cuDNN as cuDNN enables the highest performance on NVIDIA GPUs via implementation of specialized algorithms such as Winograd~\cite{winograd1980arithmetic}.

One limitation of this work is a lack of support for NVIDIA's tensor cores which is a consequence of the fact that the intermediate-level PTX assembly code~\cite{compute2010ptx} embedded within NVIDIA's cuDNN library does not include tensor core operations.  Instead, the cuDNN library appears to contain hand tuned machine-level SASS assembly code for supporting tensor cores.  This is a limitation because the current version of GPGPU-Sim only supports executing SASS code for older generation GPUs.  We believe that the updated GPGPU-Sim framework we provide is still of significant value as the limited dynamic range of 16-bit floating-point provided in NVIDIA Tensor Cores can result in convergence issues unless special steps are taken~\cite{micikevicius2017mixed}.  As a consequence, many machine learning researchers still use cuDNN APIs that avoid using Tensor Cores.  
While NVIDIA's CUTLASS~\cite{cutlass_nvidia} enables use of Tensor Cores, it does not implement highly optimized kernels such as Winograd which can provide large performance gains for convolutional neural networks (CNNs) that have small filter sizes.


Overall, we make the following contributions in this paper:

\begin{itemize}
    \item We modify GPGPU-Sim to enable running cuDNN.  In turn, this enables us to run PyTorch and should enable running other high-level frameworks such as TensorFlow.
    \item We introduce a new methodology to identify bugs in the functional simulation implementation of GPGPU-Sim.  
    \item As the runtime of architecture simulators is many orders of magnitude slower than hardware and machine learning workloads can run for days, we introduce checkpointing support to GPGPU-Sim.
    \item Using our modified GPGPU-Sim we analyze one of NVIDIA's cuDNN application samples modeling LeNet~\cite{726791} trained with the MNIST dataset.
\end{itemize}

%% file: background.tex
\vspace{-0.5ex}
\section{Background}
\label{sec:back}
\vspace{-0.5ex}

This section provides background on machine learning frameworks and their implementation as well as GPU simulators.

\subsection{Machine Learning Frameworks}
\label{subsec:back-frameworks}

Enthusiasm for employing machine learning in practice followed  AlexNet~\cite{NIPS2012_4824} achieving an impressive 15.3\% top-5 test error rate on image classification, far out-performing state-of-the-art models at that time.  AlexNet was trained for several days on two GPUs, although it was observed that the amount of GPU memory and training time limited the network's size.  
Follow on work proposed more sophisticated approaches such as VGGNet~\cite{simonyan2014very}, GoogleNet~\cite{szegedy2015going}, Residual Networks~\cite{he2016deep} and DenseNets~\cite{huang2017densely},  which have surpassed humans in classification accuracy.
This result was achieved by combining huge datasets and GPUs.  Prior research has shown that GPU can be 4$\times$ to 50$\times$ faster than  CPUs~\cite{fujimoto2008faster,lee2010debunking,mnih2009cudamat,strigl2010performance}.
Thus, GPUs play an important role in accelerating the execution times of CNNs~\cite{li2011strassen}.

Subsequently, companies have provided more optimized hardware and software for running machine learning workloads on GPUs.  For example, NVIDIA has introduced specialized cores known as Tensor Cores, high bandwidth NVLINK for communication between GPUs, and optimized their software stack (e.g., CUDA, cuBLAS, and cuDNN).  Matrix multiplication is the key underlying operation behind most of the neural network computations and a highly optimized GPU code to implement traditional matrix multiplication operation has a time complexity of $O(n^3)$~\cite{li2011strassen}.
Faster alternatives include the Winograd and Strassen~\cite{strassen1969gaussian} algorithms. Here, the Strassen's algorithm has a time-complexity of $O(n^{2.81})$ whereas Winograd has a complexity of $O(n^{2.38})$.
As Winograd is faster than Strassen's algorithm, it is used in libraries such as cuDNN, which are exploited by frameworks such as Tensorflow and PyTorch.

\subsection{GPU Performance Simulators}
\label{subsec:back-sim}

\subsubsection{NVProf}
\label{subsubsec:back-sim-prof}

The most closely related tool to GPGPU-Sim is NVProf~\cite{nvprof}, NVIDIA's command-line profiler for CUDA programs.  NVProf and GPGPU-Sim give many similar statistics, including instructions per cycle and the number of instructions executed for certain types of instructions such as loads and stores. They also track basic memory and stall information. NVProf is useful in many cases since it provides fast, accurate results from the hardware itself.  Several recent papers have used tools like 
NVProf to profile machine learning workloads~\cite{sun2018evaluating,MojumderLouis18,ZhuPhanishayee18-tbd}.  
However, since these papers use profilers, unlike our work they can 
only provide higher-level analysis about the behaviors of the 
applications.  In comparison, GPGPU-Sim provides detailed information on 
memory usage, power, efficiency, can easily be extended to provide 
additional statistics, and can output useful plots of relevant statistics 
using AerialVision~\cite{ariel2010visualizing}.

\subsubsection{Simulation}
\label{subsubsec:back-sim-sim}

Some prior work has also simulated machine learning workloads, but these 
papers used private simulators~\cite{RhuOConnor18-cdma, ParasharRhu17-scnn, HanLiu2016-eie, AlbericioDelmas2017-bitprag}.  
Since these simulators are not publicly available and few details are 
available, it is difficult to compare their approaches to ours.  In 
comparison, we simulate machine learning workloads at high fidelity in 
the widely used, publicly available GPGPU-Sim.  Moreover, the fact that 
other papers use disparate architectural simulators for machine learning 
workloads makes it crucial to provide better, publicly available tools 
for simulating machine learning workloads.

%% file: implementation.tex
\section{Implementation}
\label{sec:impl}

This section describes the modifications that were required to simulate 
cuDNN and PyTorch applications in GPGPU-Sim. 
We use regression tests and NVIDIA's cuDNN MNIST example to verify the functional correctness of our changes.  
Some of the the key changes we discuss are: (1) adding support for precompiled GPU 
kernel libraries, (2) implementing some missing CUDA Runtime and CUDA API 
functions, (3) finding and fixing bugs in the existing functional simulator that 
prevented correct execution, and (4) adding support for checkpointing simulation.  Figure~\ref{fig:cf} shows the control flow of GPGPU-Sim with our modifications.

\subsection{Support for kernels in external CUDA libraries}
\label{subsec:impl-extLibs}

\begin{figure}
  \includegraphics[width=\linewidth]{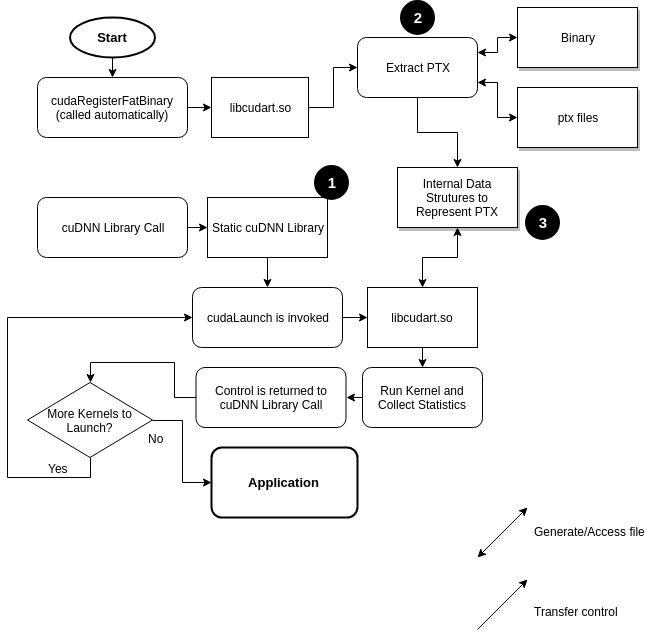}
  \caption{GPGPU-Sim's Control Flow with modifications for cuDNN.}
  \label{fig:cf}
\end{figure}

The existing version of GPGPU-Sim first extracts all PTX code embedded within an application from the binary using an NVIDIA supplied program called {\tt cuobjdump}.  Next, GPGPU-Sim combines the extracted PTX into a single PTX file that is then parsed by GPGPU-Sim's program loader.  Unfortunately, this approach causes two issues when trying to run cuDNN enabled applications:

First, we found that cuDNN programs and programming frameworks that use cuDNN are typically dynamically linked to the cuDNN library.  However, current versions of {\tt cuobjdump} do not resolve dynamic linked libraries before searching for PTX code.  Thus, GPGPU-Sim fails to launch kernels contained in dynamically linked libraries.  There are two potential solutions to this: either modify GPGPU-Sim to search through any dynamically linked libraries (using {\tt ldd}), or rebuild the CUDA application and statically link against the external library.  In this paper, we followed the latter approach, as shown in Figure~\ref{fig:cf} (\mycirc{1}).

Second, cuDNN includes code and variables with the same names in multiple source files.  After combining all the PTX extracted from the application binary into a single file, these multiple definitions resulted in errors when parsed by GPGPU-Sim's program loader.  Thus, we modified GPGPU-Sim to extract and process each embedded PTX file separately (\mycirc{2}).   

\subsection{Additional CUDA Language Support}
\label{subsec:impl-cuda}

NVIDIA's CUDA enables overlapping memory copies from CPU to GPU with computation on the GPU via a construct known as streams (similar to a command queue in OpenCL).  We found that cuDNN uses multiple streams to overlap memory transfers with computation.  Although GPGPU-Sim already supports streams, we found it did not support all the required API functions.  Thus, we added support for {\tt cudaStreamWaitEvent}, an API call that allows a stream to wait for an event to occur before continuing execution.

Additionally, we also added support for PTX instructions that were not implemented in the current version of GPGPU-Sim, but are used by cuDNN.  For example, we found that cuDNN uses the bit reverse instruction ({\tt brev.type d, a;}), which was introduced in PTX version 2.0, for FFT-based convolutional kernels~\cite{chetlur2014cudnn, fftcnn}.  Thus, we added an implementation for this instruction, which is used to output the bits of its input in reverse order.

Moreover, in the process of developing our debugging tool (discussed further in Section~\ref{subsec:impl-debug}), we found that we had to add an alternative CUDA API call for launching kernels: {\tt cuLaunchKernel}.  The CUDA Runtime API equivalent is {\tt cudaLaunch}, which was already supported by GPGPU-Sim.

\subsection{Texture References}
\label{subsec:impl-textures}

To represent textures, GPGPU-Sim uses a system of texture names, texture references (texref), cudaArrays, textureInfos, and textureReferenceAttrs. A given texture name maps to a texture reference, and a given texture reference maps to a set of cudaArray, textureInfo, and textureReferenceAttr. This aligns with the APIs \_\_cudaRegisterTexture to map a name to a texref; cudaBindTextureToArray to map a texref to a set of cudaArray, textureInfo, and textureReferenceAttr; and unbindTexture to unbind a cudaArray from a texref. A texture instruction in CUDA kernels accesses this data by looking it up with the texture name. 

Although textures were already supported in GPGPU-Sim, MNIST registered multiple texrefs to the same name.  This caused conflicts in the map and as a result, data was lost.  Consequently, some texture instructions would fail because they could not find the cudaArray they were looking for.  To resolve  this problem, we mapped the texture names to a set of texrefs and also mapped texture names directly to their associated cudaArrays, textureInfos, and textureReferenceAttrs. Thus, texture instructions now use texture name to look up cudaArrays, textureInfos, and textureReferenceAttrs. 

We also encountered another problem with textures where the program called bindTextureToArray on the same texref with different cudaArrays multiple times. To resolve this, we assume the program meant to first unbind the existing cudaArray from the texref, and then bind the new cudaArray to the texref. 

\begin{center}
  \begin{figure*}
    \centering
\includegraphics[width=0.35\linewidth]{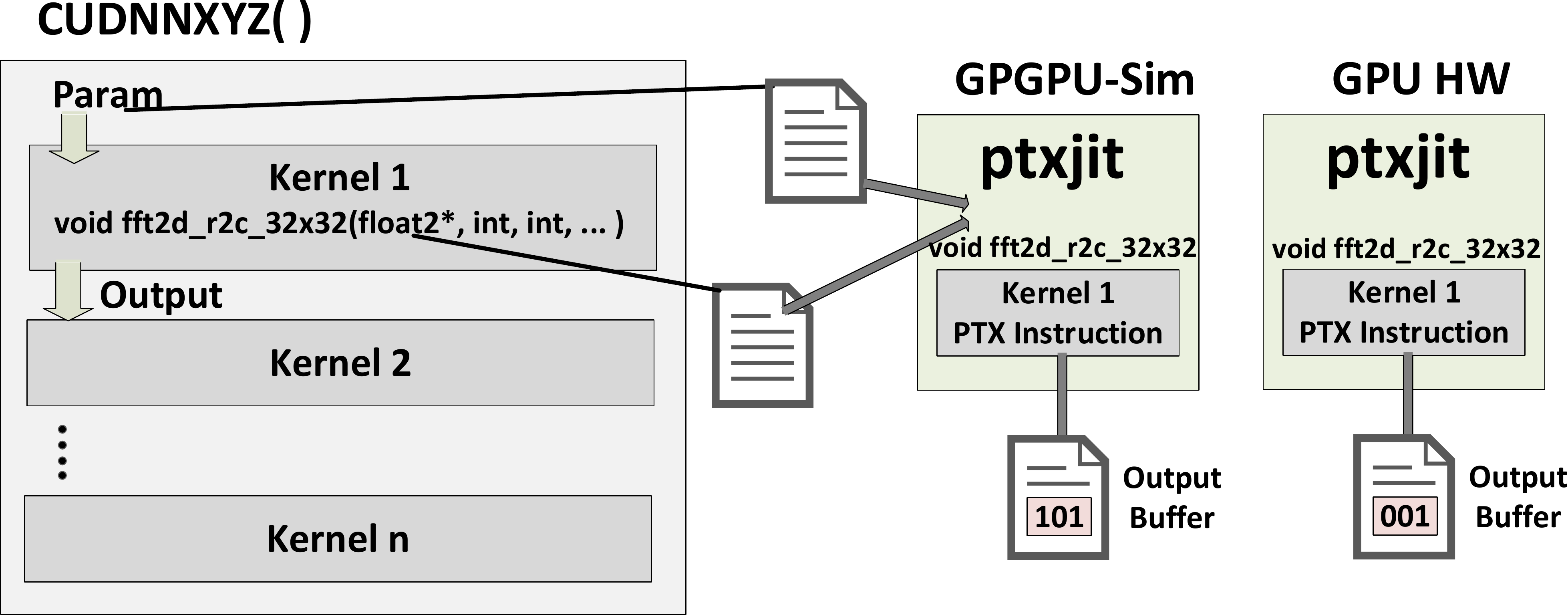}
    \caption{Identifying the first incorrectly executing kernel within a multi-kernel library function call}
    \label{fig:cp2}
  \end{figure*}    
\end{center}

\begin{center}
  \begin{figure*}
    \centering
    \includegraphics[width=0.85\linewidth]{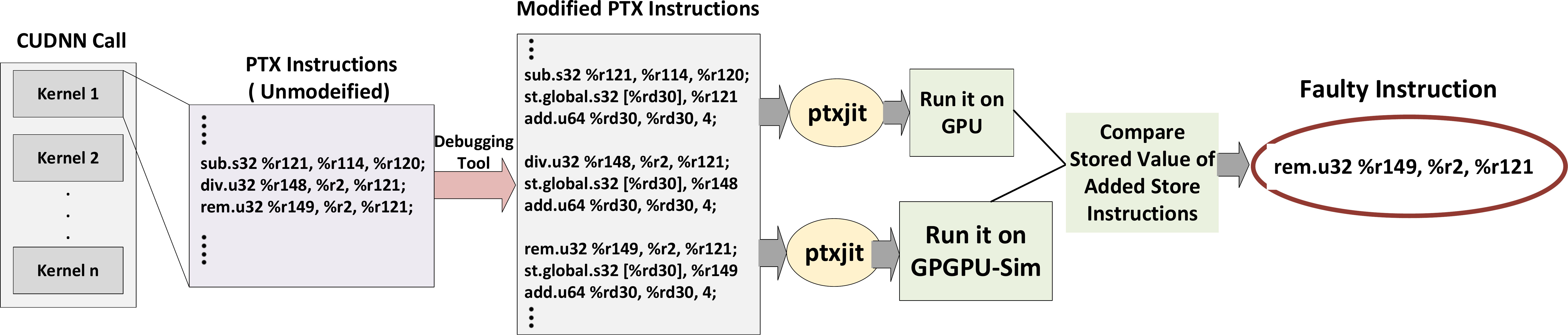}
    \caption{Identifying the first incorrectly executing instruction within the first incorrectly executing kernel}
    \label{fig:cp3}
  \end{figure*}    
\end{center}
\vspace{-1cm}

\subsection{Debugging Functional Simulation Errors}  
\label{subsec:impl-debug}

Although the existing GPGPU-Sim simulator correctly runs a large number of CUDA applications such as those found in Rodinia~\cite{Che:2009:RBS:1678998.1680782},  machine learning workloads presented additional complexities.  After the changes discussed in Sections~\ref{subsec:impl-extLibs}-\ref{subsec:impl-textures}, GPGPU-Sim could run MNIST to completion but generated incorrect results.  This presented us with a big challenge:  How to find which instruction(s) out of billions of executed instructions was incorrect?  

The developers of GPGPU-Sim gave us some help by explaining the process they followed to getting GPGPU-Sim working~\cite{bakhoda2009analyzing}: first, they validated individual instructions one-by-one by comparing execution on real GPU hardware with execution on GPGPU-Sim.  Then, once each instruction appeared to work, they started with getting smaller applications (e.g., {\tt template} from the CUDA SDK) running and progressively ran larger applications.  While adding applications, they used GNU's {\tt gcov} tool to compare the coverage analysis of the simulator for correctly simulated applications and new, incorrectly simulated applications.  By comparing these, performing ``differential coverage analysis'', they were able to narrow down which part of the functional simulator had a bug.

Thus, we first attempted to solve the functional correctness issues by employing a similar differential coverage analysis.  We compared the coverage of the functional simulator when running the regression tests on GPGPU-Sim's Github page with the results obtained running MNIST.  Using this approach, we identified that GPGPU-Sim's implementation of the bit field extract instruction ({\tt bfe.type d, a, b, c;}) had subtle errors for signed inputs.  Thus, we modified GPGPU-Sim's bit field extract instruction implementation to correctly handle signed 32-bit and 64-bit integer inputs.  However, after this change, GPGPU-Sim's result for MNIST was still incorrect.  We found no other other lines that were exercised by cuDNN and not exercised by the regression tests that appeared to involve incorrectly executing instructions.   

We then developed a new approach to debugging functional simulation errors in GPGPU-Sim that was ultimately successful in finding the remaining error.  We believe this approach and the resulting debug tool, which we plan to make available with this paper, will be useful to other researchers encountering incorrect results when running new applications in GPGPU-Sim.  At a high level, we compare the execution of every instruction executed by GPGPU-Sim to the result obtained from executing that instruction on hardware, then flag the first instruction with an error.  An important practical complication we encountered is that every high level API call in cuDNN launches several kernels onto the GPU.  Thus, in practice, we followed a three-step process: first identify which cuDNN API call results in incorrect results, then identify which GPU kernel launched within that API call is executing incorrectly, and finally identify the first instruction in that kernel that executed incorrectly.  

To identify which cuDNN API call was incorrect, we compare the result buffers on GPGPU-Sim versus the hardware by adding calls to {\tt cudaMemcpy} to MNIST.  However, having identified an incorrect API call, finding which specific kernel of our cuDNN enabled application was responsible for an incorrect result is non-trivial because we do not have source code for cuDNN.  Thus, we changed to GPGPU-Sim
to optionally capture and save all relevant data to a file.  As shown in Figure~\ref{fig:cp2}, this data corresponds to the data which is being copied to the GPU before a kernel is launched, along with the parameters passed into the kernel as GPGPU-Sim runs.  

Armed with this data, and using our debugging framework, the extracted PTX, and a version of the {\tt ptxjit} CUDA SDK example, we systematically launch each kernel from the failing cuDNN API call onto both GPGPU-Sim and a real GPU.  We assume that any kernel parameter that is a pointer may point to an output buffer.  We also modified GPGPU-Sim to obtain the size of any GPU memory buffers pointed to by these pointers.  Then, after the extracted kernel executes 
we use {\tt cudaMemcpy} to transfer all buffers back to the CPU so they can be output to a log and compared.  By comparing the buffers after each kernel, we can identify which kernel executed incorrectly.

Next, as illustrated in Figure~\ref{fig:cp3} we instrumented the extracted PTX for just the incorrectly executing kernel so that the results of each executed instruction that writes a value to a register is saved into a new global array in GPU memory. At the end of the kernel execution, this array is transferred to CPU memory and written to a log file.  Comparing GPGPU-Sim execution of this modified kernel versus GPU hardware helps the user to identify the first instruction that executed incorrectly. To help automate the process of adding store instructions to a kernel, we developed an LLVM-based tool to modify a kernel.

Using the above approach we found that the first kernel in 
{\tt cudnnConvolutionForward} had an error when executing a remainder 
instruction.  Specifically, the remainder instruction 
``{\tt rem.u32 \%r149, \%r2, \%r121;}'' 
inside the kernel ``\lstinline{fft2d_r2c_32x32}'' generated a different result in GPGPU-Sim versus hardware.  In the existing GPGPU-Sim this instruction is implemented by the function \lstinline{rem_impl} using the code:
\begin{lstlisting}
   data.u64 = src1_data.u64 % src2_data.u64;
\end{lstlisting}

In GPGPU-Sim, \lstinline{data}, \lstinline{src1_data}, and \lstinline{src2_data} are C/C++ ``union'' type called \lstinline{ptx_reg_t} which, among others, contains fields named ``\lstinline{.u32}'' for holding 32-bit unsigned values and ``\lstinline{.u64}'' for unsigned 64-bit values.  Thus, this code is incorrect in some cases because it does not take into account signed vs. unsigned operations and 32- vs. 64-bit values.  To resolve the problem, we added a switch statement take account of the type specifier and signed operations, e.g., {\tt .u32} and instead use the code such as:
\begin{lstlisting}
   data.u32 = src1_data.u32 % src2_data.u32;
\end{lstlisting}
when type is ``\lstinline{.u32}'' and 
\begin{lstlisting}
   data.s32 = src1_data.s32 % src2_data.s32;
\end{lstlisting}
when type is ``\lstinline{.s32}''.  After making this change, cuDNN was able to run 32-bit floating-point applications correctly.

\subsubsection{FP16 Support}
\label{subsubsec:impl-debug-fp16}

Much research on hardware support for deep learning, particularly for inference, 
focuses on reduced precision.  NVIDIA supports 16-bit floating-point on both their 
regular ALUs and in their Tensor Cores and cuDNN has 16-bit (FP16) versions of the 
algorithms it supports for deep learning.  Accordingly, we added FP16 support in GPGPU-Sim, including instructions that convert FP32 to FP16 and back 
using an open source library.  

However, when we ran MNIST in FP16 mode, it produced incorrect results.  We traced this problem back to a subtle issue with multiply instructions, followed by either 
a subtract or an add, 
being optimized by the NVIDIA assembler into fused-multiply-add (FMA) SASS 
instructions.  The FMA instruction retains additional precision between 
multiplication and addition, which results in a mismatch between GPGPU-Sim and 
execution on GPU hardware.  Thus, correctly simulating code with 16-bit 
floating-point instructions is left to future work.  We expect our 
debugging strategy mentioned above will be of help in this regard but will need 
to be modified to account for rounding errors.

\subsubsection{Timing-Model Deadlocks}
\label{subsubsec:impl-debug-timing}

We also fixed bugs in the memory model and in GPUWattch code that caused cuDNN enabled programs to deadlock GPGPU-Sim's timing model.

\subsection{PyTorch and TensorFlow}
\label{subsec:impl-pyTorch_TF}

After successfully running 32-bit MNIST with correct outputs, we turned our attention to supporting PyTorch and TensorFlow. PyTorch's calls invoke functions in its shared library, {\tt \_C.so}. A regular PyTorch build uses rpath to link this {\tt \_C.so} dynamically to a hard-coded path to the CUDA Runtime Library installed on the machine. We removed all these rpath links, so that it would forced to look for a shared library at run time. Then we changed the search paths in our environment so that it would find GPGPU-Sim's {\tt libcudart.so}. Finally, we used cuDNN's shared library {\tt libcudnn.so} to get the source of the corresponding PTX.

When an application imports Torch, the library {\tt libcudart.so} is loaded, which invokes a series of initialization functions in GPGPU-Sim and GPUWattch. A thread\_exit in GPUWattch caused another library load, which created a deadlock. We solved this by removing the thread\_exit.   

We took a similar approach to try to get Tensorflow to run in GPGPU-Sim.  We managed to get TensorFlow to call CUDA Runtime API, but unfortunately TensorFlow tries to launch PTX that is not in {\tt libcudnn.so}.  To get around this, we attempted to use TensorFlow's {\tt \_pywrap\_tensorflow\_internal.so}'s PTX.  However, this file it uses syntax that is not supported by GPGPU-Sim to initialize arrays using curly braces (\{\}). Thus, adding this support is left to future work.

\subsection{Checkpointing support}
\label{subsec:impl-ckpt}

\begin{figure}[h!]
  \includegraphics[width=\linewidth]{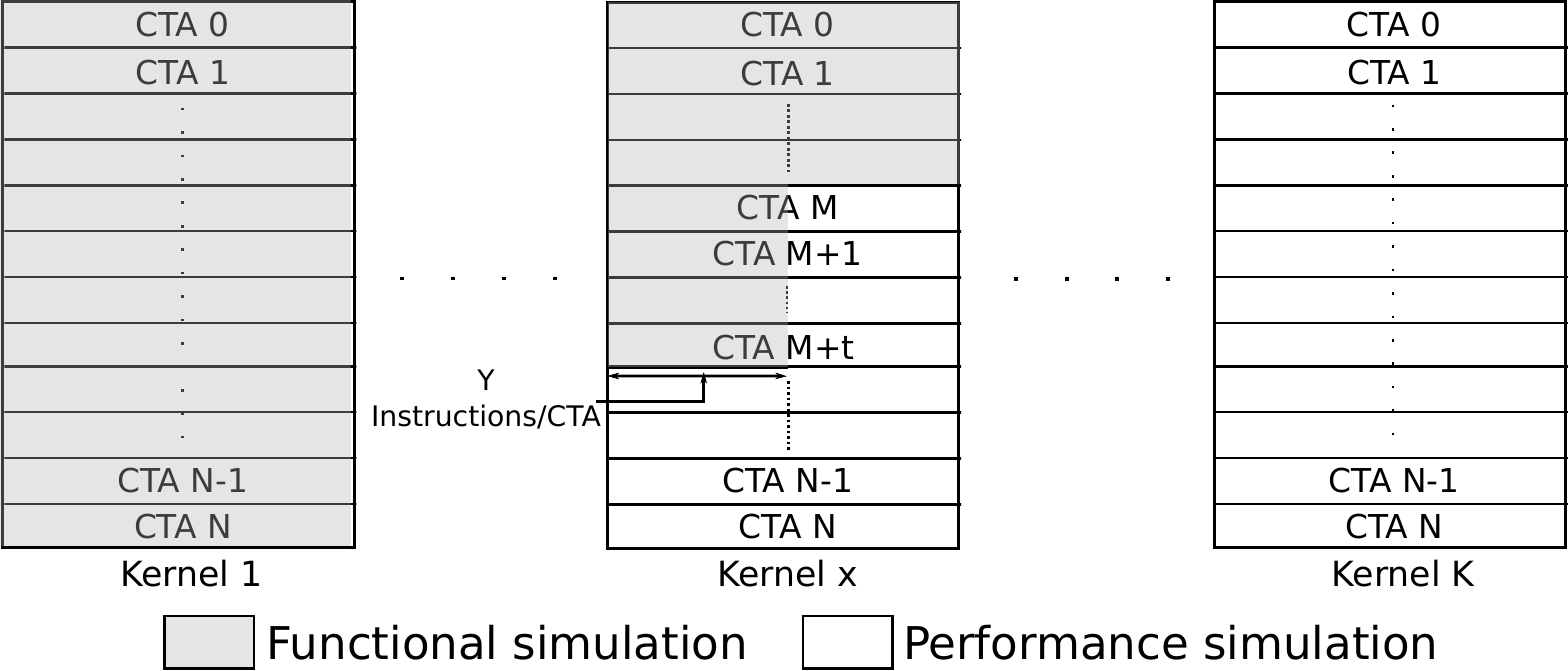}
  \caption{Checkpointing in GPGPU-Sim: Instead of running entire the application in the Performance mode, the user can run the application in the Functional simulation mode until some point, as shown in Figure~\ref{fig:cp}, and save the necessary data to resume in files. Then the user can resume from this point in Performance simulation mode.  Parameters to define checkpoint position such as $x$, $M$, $t$ and $y$, as shown in Figure~\ref{fig:cp}, can be defined by the user in GPGPU-Sim config file.}
  \label{fig:cp}
\end{figure}

\begin{figure*}[t!]
  \centering
  \includegraphics[width=0.80\linewidth]{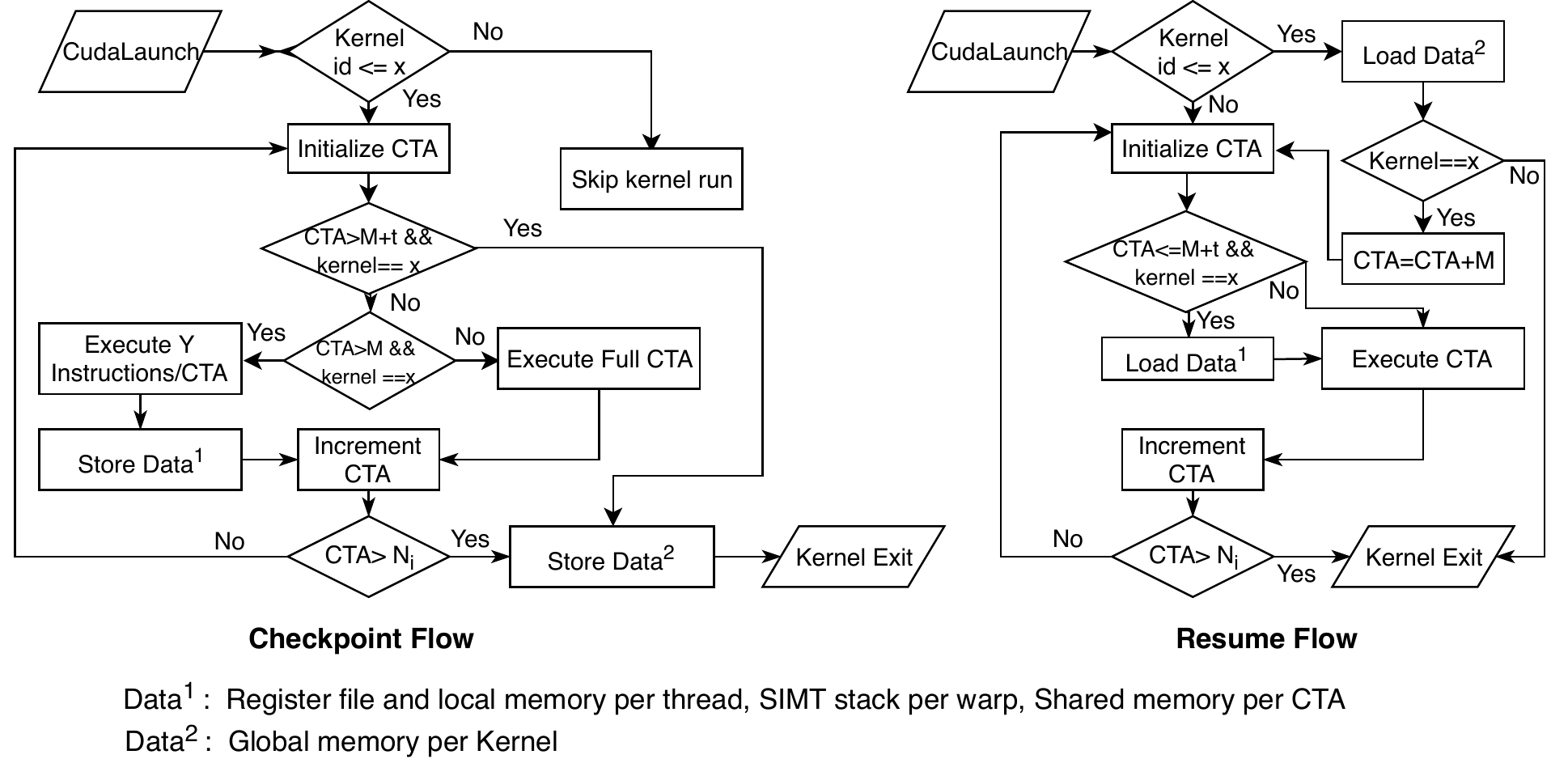}
  \caption{Block diagram for Checkpoint and resume flow. Here parameters x, t, M and Y can be configured by the user and explained in figure \ref{fig:cp}. $N_i$ is the total number of CTA in the specific kernel. }
  \label{fig:cp_block}
\end{figure*}

GPGPU-Sim can be run either in the Performance simulation mode or in the Functional simulation mode.  The Functional simulation mode only executes the application and gives functionality correctness of the application, but doesn't give any performance statistics. The Performance simulation mode collects all statistics and gives an estimate of the number of GPU clock cycle on actual hardware. However, the Performance simulation mode is generally 7-8 times slower than the Functional simulation mode. Run-time of architecture simulator is significantly slower than the actual hardware and accordingly a typical cuDNN/Pytorch application might take very long time to run on GPGPU-Sim.  For example, MNIST takes $\sim $1.25 hours on GPGPU-Sim's Performance mode to classify three images. This is much longer than a real GPU takes, and the user may only be interested in the performance analysis of a particular part of the program rather than the entire program.  Thus, we added checkpoint-resume functions to GPGPU-Sim. The flow of checkpointing is explained in the Figure~\ref{fig:cp} and the flow of the implementation is explained in the Figure~\ref{fig:cp_block}. 

We support checkpointing at bound kernel boundaries and a CTA boundary within a kernel.  At the end of each kernel, we save the computational results to the GPU's global memory.  This makes it possible to resume execution from any kernel which has been executed before the checkpoint.  For a checkpoint at a specific CTA $M$ within a kernel $x$, all the kernels with kernel\_id $< x$ are executed normally and the state of the GPU's global memory is saved to a file. For kernel $x$ any of the $M-1$ CTAs before the desired checkpoint point are executed normally. However, for CTAs $M$ to $M+t$, $y$ instructions ($y > x$) per CTA are executed.  In order to resume from that point, we also need to checkpoint the register data and local memory data for each active thread, the SIMT stack (which is used to handle branch divergence within a warp~\cite{aamodt2012gpgpu}) for each active warp, and the shared memory for each CTA.  All kernels with kernel\_id $> x$ are not executed.

To resume at a given checkpoint, all kernels with a kernel\_id $< x$ are skipped but the GPU global memory is restored for each kernel since the program might call {\tt cudaMemcpy} between two kernels and perform computation on this data. For kernel $x$, all CTAs $< M$ are skipped for computation.  CTAs $M$ to $M+t$ are initialized and the register data, local memory, SIMT stack, and shared memory are restored for the corresponding threads, warps and CTAs.   

%% file: correlation.tex
\vspace{-1ex}
\section{Correlation}
\label{sec:correlation}
\vspace{-1ex}

During the process of updating GPGPU-Sim to run machine learning workloads 
(Section~\ref{sec:impl}), we also used a 32-bit floating-point version of MNIST 
to correlate GPGPU-Sim's execution time with a GeForce GTX 1050.  We use MNIST to 
perform the correlation because it is relatively simple and uses a wide variety 
of cuDNN layers such as LRN and Winograd.  Additionally, MNIST contains 
self-checking code at the end of the application, which helps ensure the functional 
correctness of our implementation.

\begin{figure}[tb]
  \begin{center}
    \includegraphics[width=\linewidth]{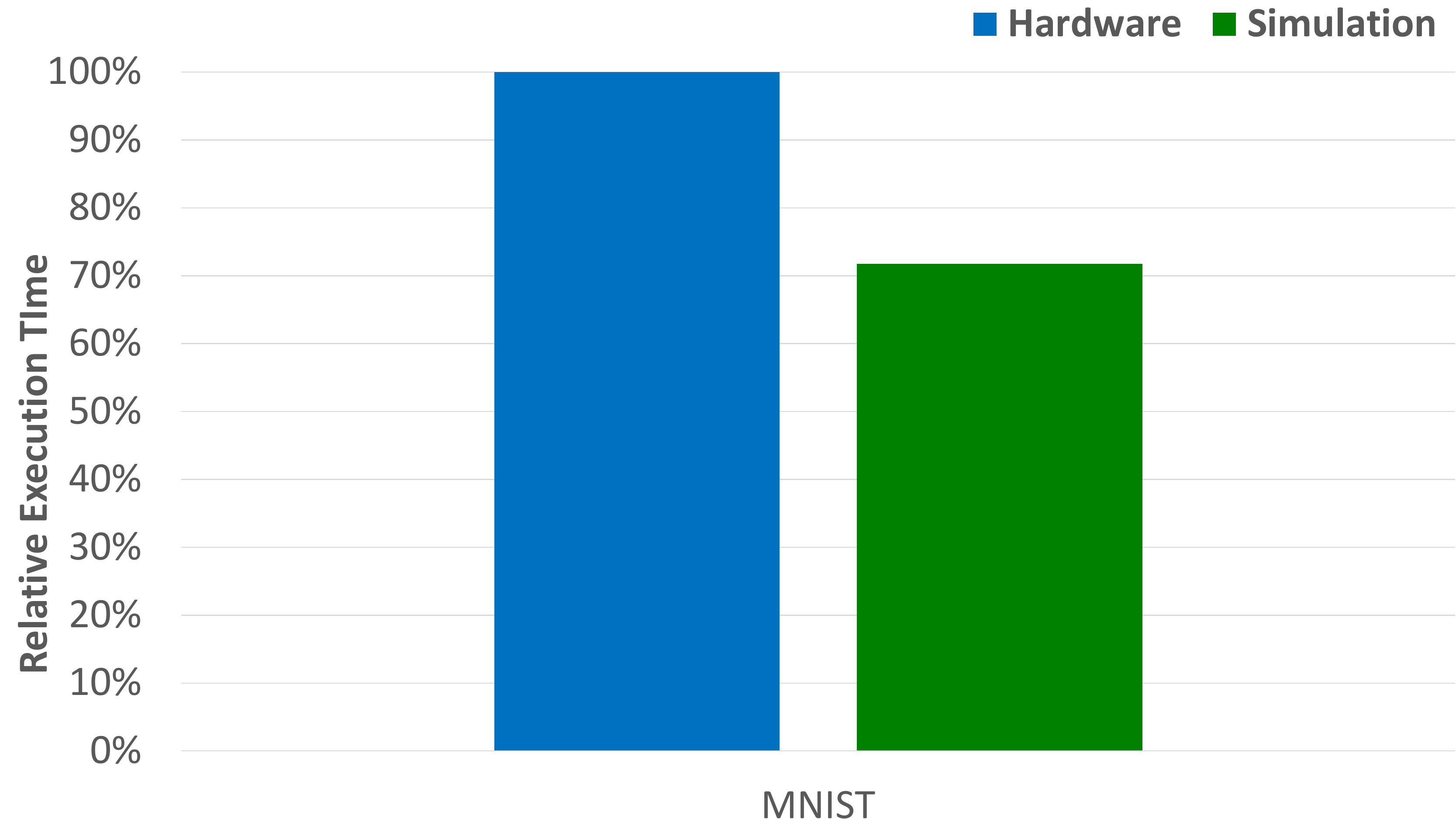}
    \caption{Correlating execution time for MNIST.}
    \label{fig:mnist-cycles-rel}
  \end{center}
\end{figure}

\begin{figure}[tb]
  \begin{center}
    \includegraphics[width=\linewidth]{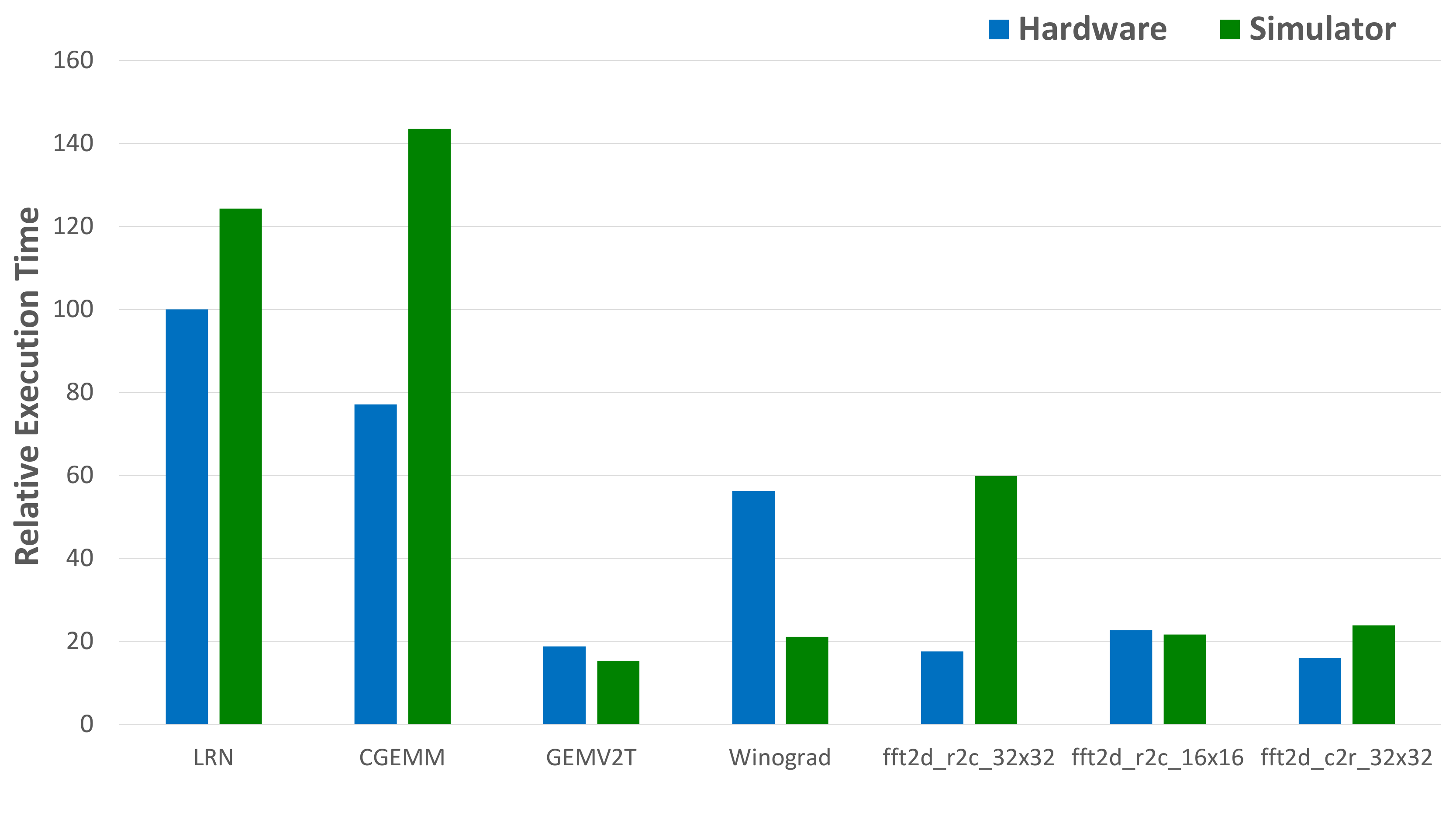}
    \caption{Select per-kernel, execution time correlation results for MNIST.  These results are a subset of the kernels in Figure~\ref{fig:mnist-cycles-rel}.}
    \label{fig:mnist-cycles-perKernel-rel}
  \end{center}
\end{figure}

We correlated GPGPU-Sim's performance with real GPUs by comparing the number of GPU 
cycles with those reported by NVProf.  Figures~\ref{fig:mnist-cycles-rel} and 
\ref{fig:mnist-cycles-perKernel-rel} show the overall correlation for MNIST and 
the correlation for select kernels, respectively.  We selected these kernels 
because they show the largest discrepancies.  For the kernels we do not show, 
GPGPU-Sim achieves very high correlation with the real GPU.  Overall, GPGPU-Sim 
provides a correlation of 72\%.  Inspecting the per-kernel results shows that the 
overall discrepancy is heavily affected by a few kernels such as CGEMM, Winograd, 
and LRN.  Thus, improving support for these kernels will make the overall 
correlation even better.

\subsection{Power Consumption}
\label{subsec:correlation-power}

\begin{figure}[tb]
    \begin{center}
        \includegraphics[width=\linewidth]{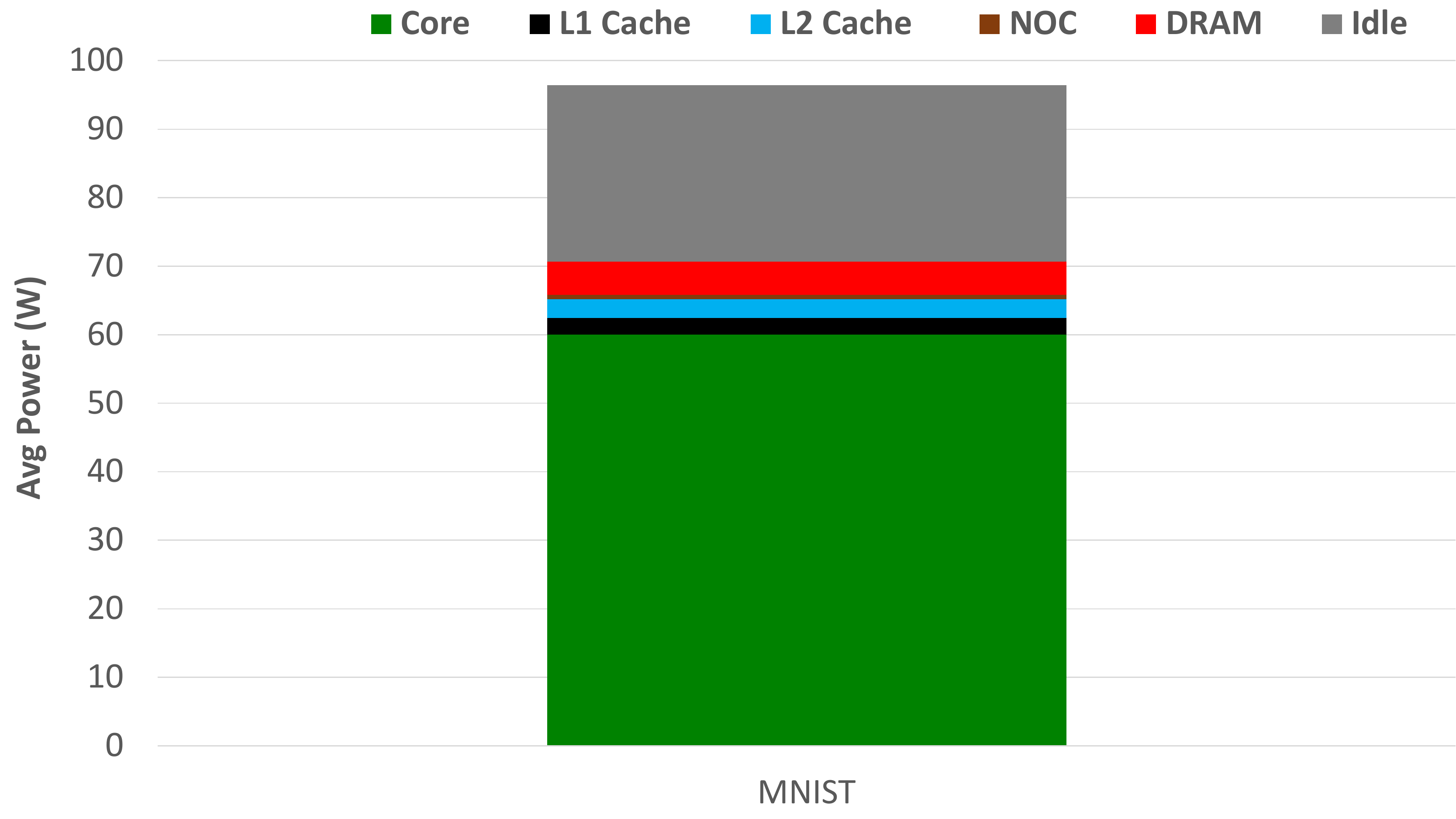}
        \caption{Average power consumption for a 32-bit floating-point version of MNIST, subdivided into 6 components of the simulated GPU.}
        \label{fig:mnist-power}
    \end{center}
\end{figure}

Figure~\ref{fig:mnist-power} breaks down MNIST's power consumption 
into 6 key categories: core, L1 cache, L2 cache, NOC, DRAM, and 
Idle.  As expected for relatively computationally intensive CNNs 
like MNIST, on average the core (in particular the ALUs) consume 
65\% of the power.  However, on average Idle power consumes a further 25\% of the total power.  This represents a tuning opportunity for 
future architectural exploration, which is enabled by this work.

%% file: caseStudy.tex
\vspace{-1ex}
\section{Case Studies}
\label{sec:cases}
\vspace{-1ex}

\subsection{Methodology}
\label{subsec:cases-meth}

In this section, we study another simple cuDNN program from the NVIDIA examples, conv\_sample.  We choose conv\_sample because it performs forward, backward data, and backward filter convolutions, which are common machine learning operations. Using conv\_sample, we iterated over the various cuDNN algorithms available for each type of convolution. For forward convolution, we ran FFT, FFT Tiling, GEMM, Implicit GEMM,  Winograd, and Winograd Nonfused. For backward data convolution, we ran Algorithm 0, Algorithm 1, FFT Tiling, Winograd, and Winograd Nonfused. For backward filter convolution, we ran Algorithm 0, Algorithm 1, Algorithm 3, FFT, FFT Tiling, and Winograd Nonfused.  For all of these different approaches, we model a NVIDIA Pascal GeForce GTX1080Ti in GPGPU-Sim.  

We studied each algorithm using AerialVision~\cite{ariel2010visualizing}, a tool that plots metrics per bank/shader per cycle.  In this case study, we plotted DRAM efficiency, global IPC and per shader IPC, and warp divergence -- details that are enabled by using simulators like GPGPU-Sim. 
DRAM efficiency and utilization is DRAM bandwidth utilization when there is a pending request waiting to be processed and two times the number of read and write commands per command cycle~\cite{aamodt2012gpgpu}. The y-axis for each is the bank number. 
The global IPC shows the total number of instructions committed per cycle from all the shader streaming multiprocessors, or cores. The shader IPC breaks this down further and shows the number of instructions being committed per shader core. To show how IPC varies across shaders, in the graph the y-axis is the shader core number. 
Warp divergence is plotted as a breakdown of the number of warps that are issued for execution. W0 means idle, and W1 through W32 are warps. Visually, the more layers there are, the more warp divergence there is. 

\begin{figure}
  \includegraphics[width=\linewidth]{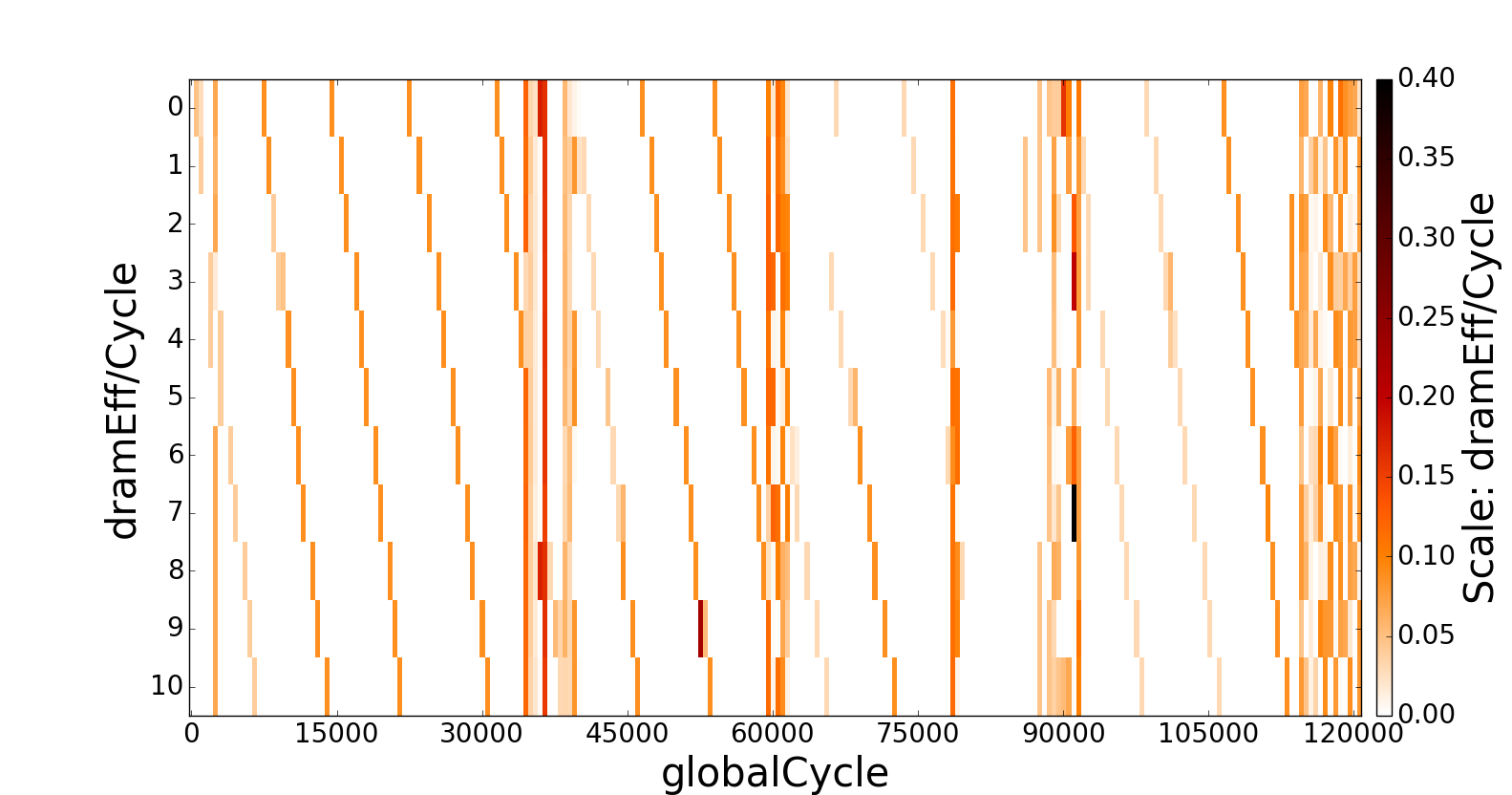}
  \caption{Forward Convolution (FFT) DRAM Efficiency Plot}
  \label{fig:dramEff_fft}
\end{figure}

\begin{figure}
  \includegraphics[width=\linewidth]{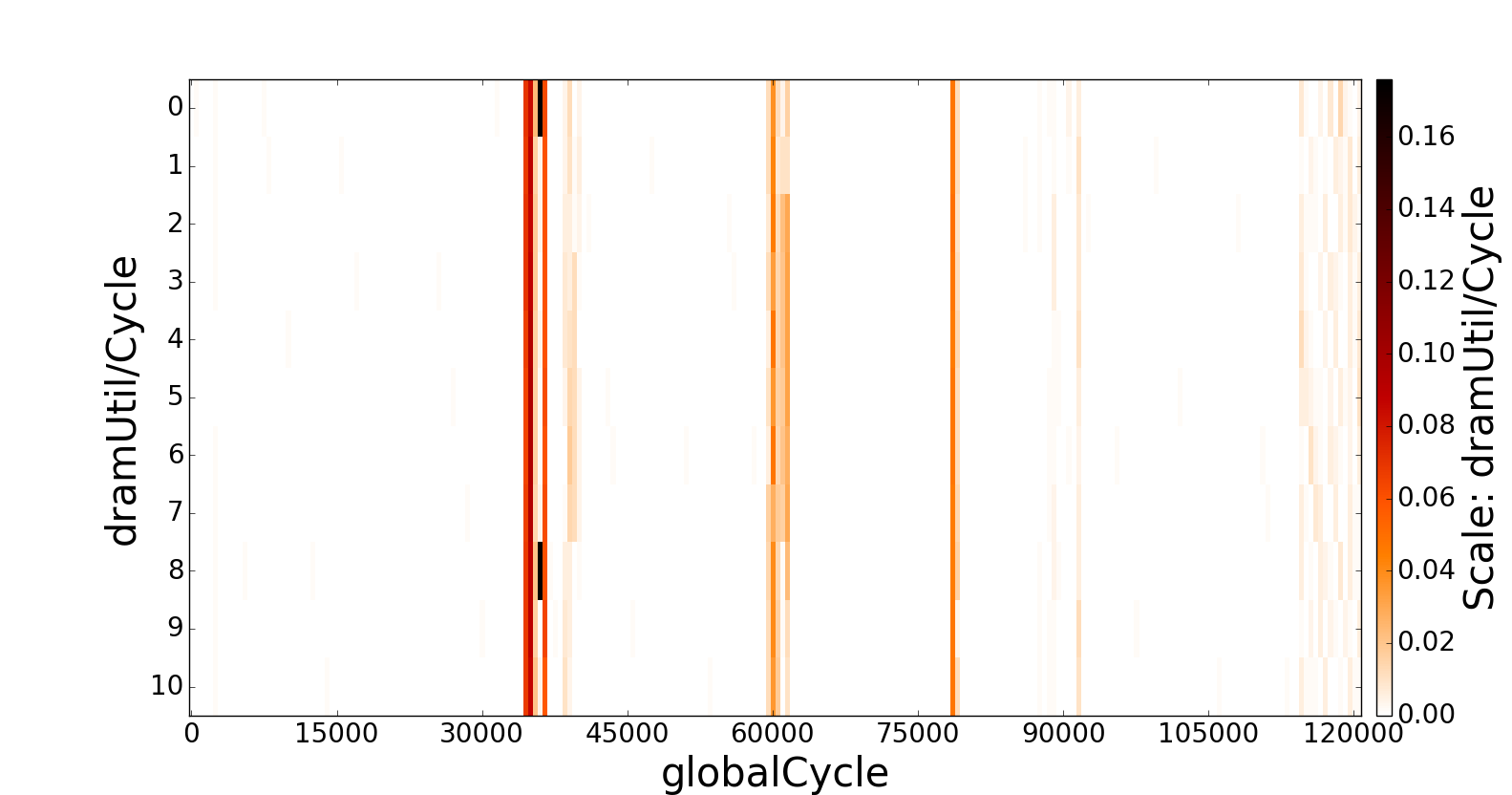}
  \caption{Forward Convolution (FFT) DRAM Utilization Plot}
  \label{fig:dramUtil_fft}
\end{figure}

\begin{figure}
  \includegraphics[width=\linewidth]{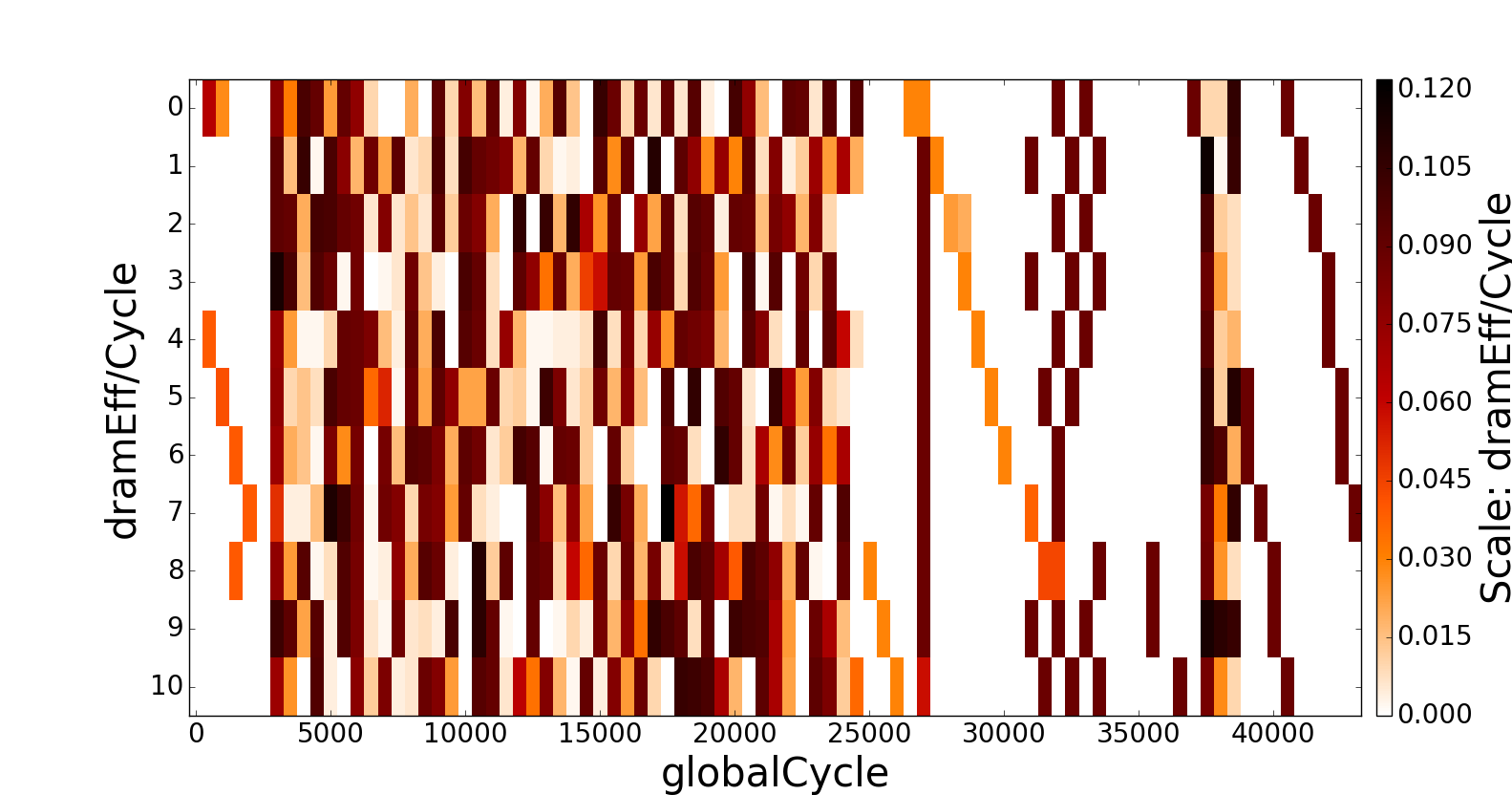}
  \caption{Forward Convolution (GEMM) DRAM Efficiency Plot}
  \label{fig:dramEff_gemm}
\end{figure}

\begin{figure}
  \includegraphics[width=\linewidth]{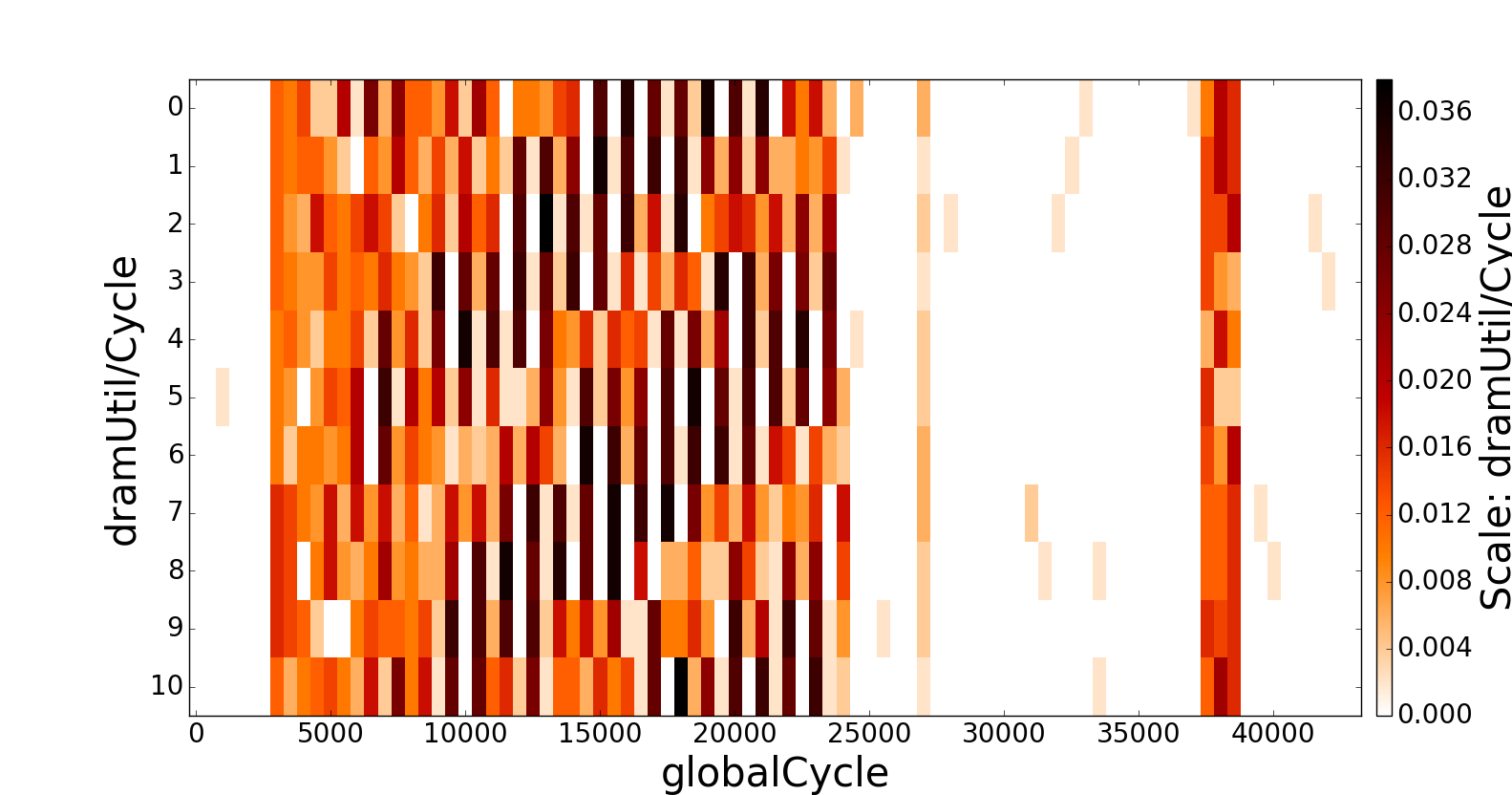}
  \caption{Forward Convolution (GEMM) DRAM Utilization Plot}
  \label{fig:dramUtil_gemm}
\end{figure}

\begin{figure}
  \includegraphics[width=\linewidth]{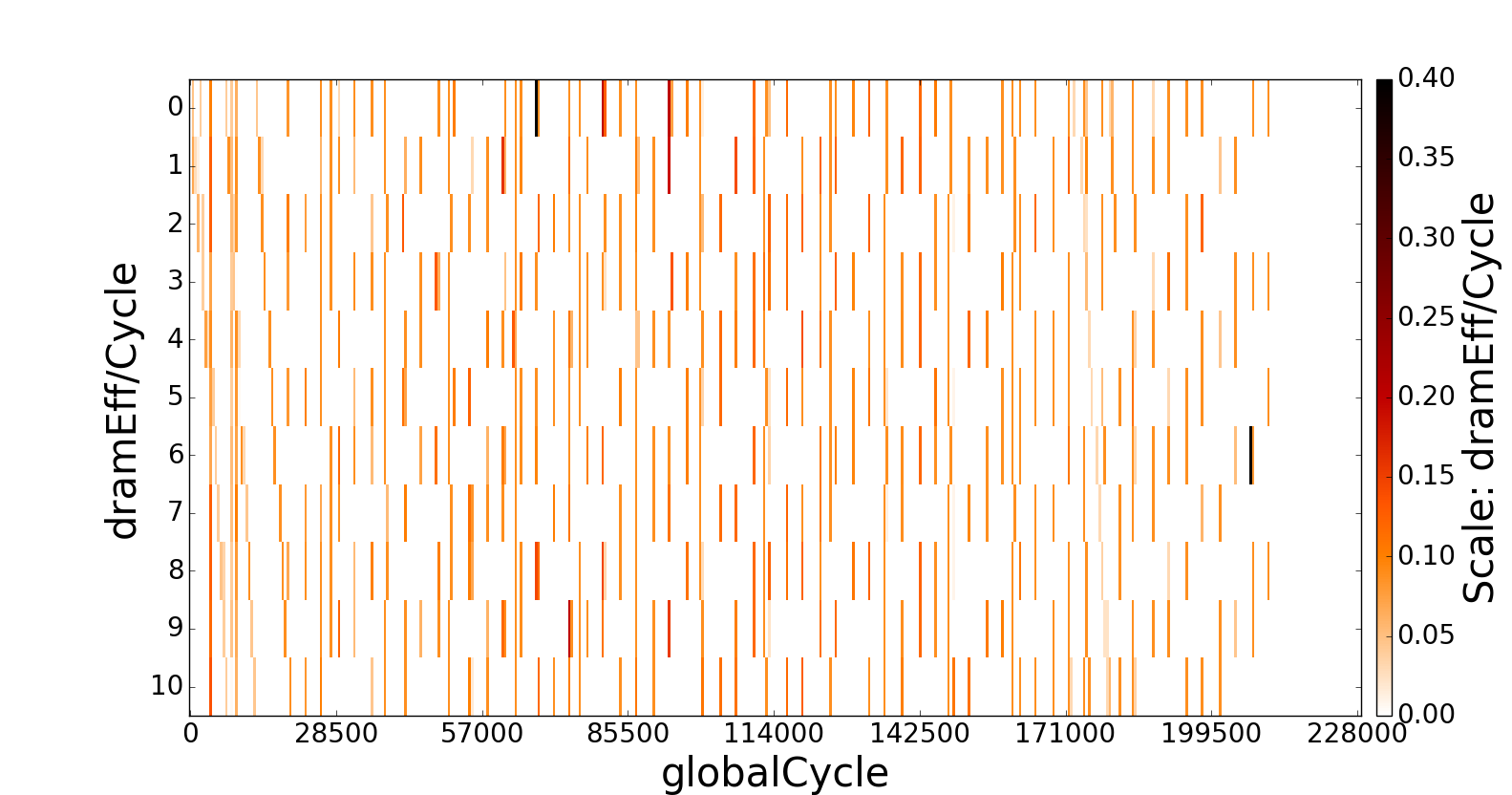}
  \caption{Backward Filter Convolution (Algorithm 0) DRAM Efficiency Plot}
  \label{fig:dramEff_0}
\end{figure}

\begin{figure}
  \includegraphics[width=\linewidth]{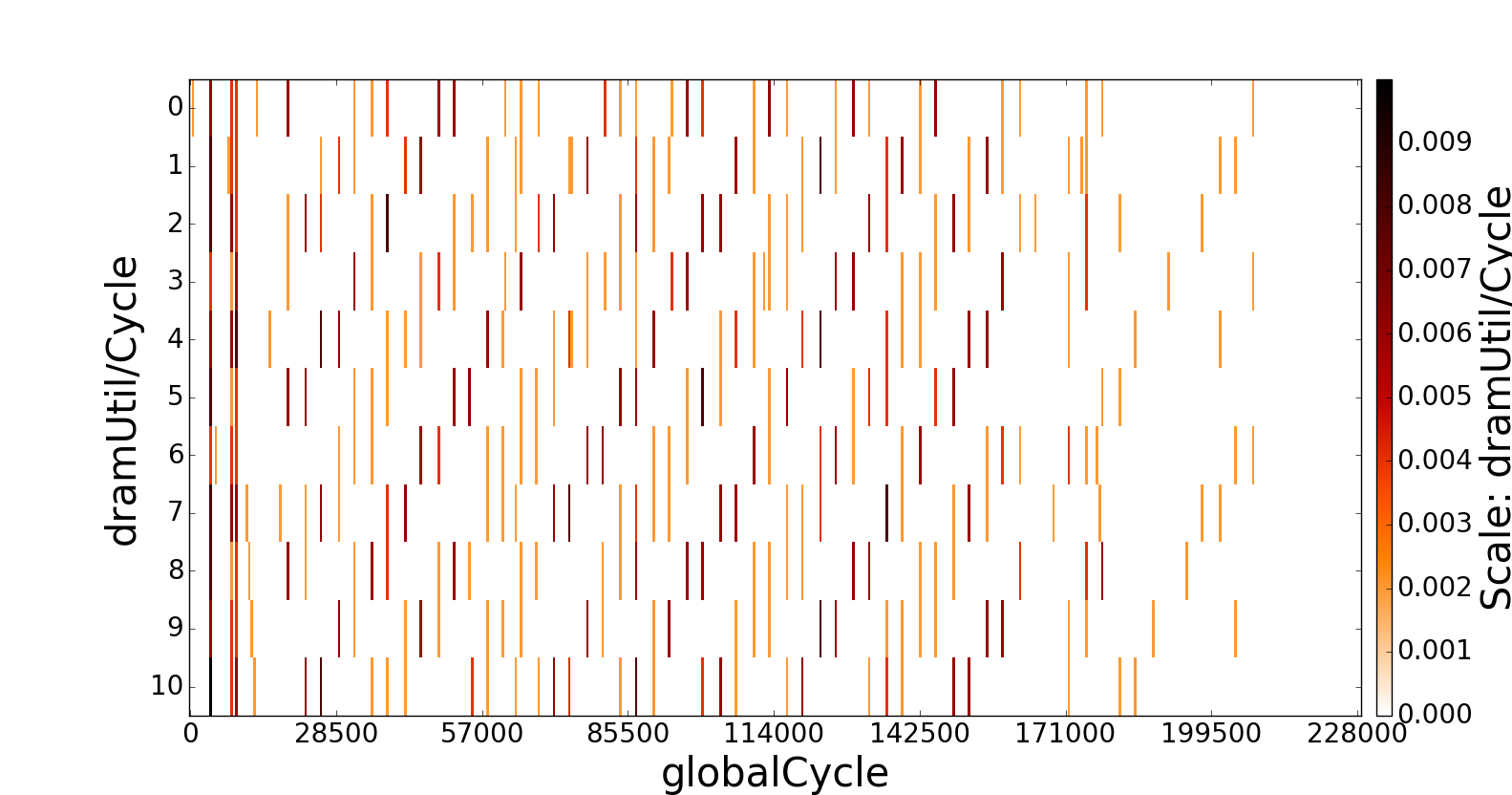}
  \caption{Backward Filter Convolution (Algorithm 0) DRAM Utilization Plot}
  \label{fig:dramUtil_0}
\end{figure}

\subsection{DRAM Efficiency and Utilization}
\label{subsec:cases-dram}

Figure~\ref{fig:dramEff_fft}, \ref{fig:dramEff_gemm}, \ref{fig:dramUtil_gemm}, \ref{fig:dramEff_0}, and \ref{fig:dramUtil_0} shows the DRAM efficiency and utilization, for each convolution approach, as a sequence of DRAM banks.  For FFT, we see that most of the DRAM banks show high memory efficiency, interspersed with periods of parallel efficiency.  However, FFT also has a mix of serial and parallel efficiency patterns.  In the serial sections, FFT is unable able to parallelize memory bank accesses. This phenomenon is known as bank camping. However, bank camping is less of an issue for other approaches like forward convolution with the GEMM algorithm and the backward filter convolution with either algorithm 0 or 1.  More generally, some algorithms can make more efficient use of all the banks at the same time, and for all approaches the memory access patterns tend to go through phases of high and low efficiency.

\begin{figure}[tb!]
  \includegraphics[width=\linewidth]{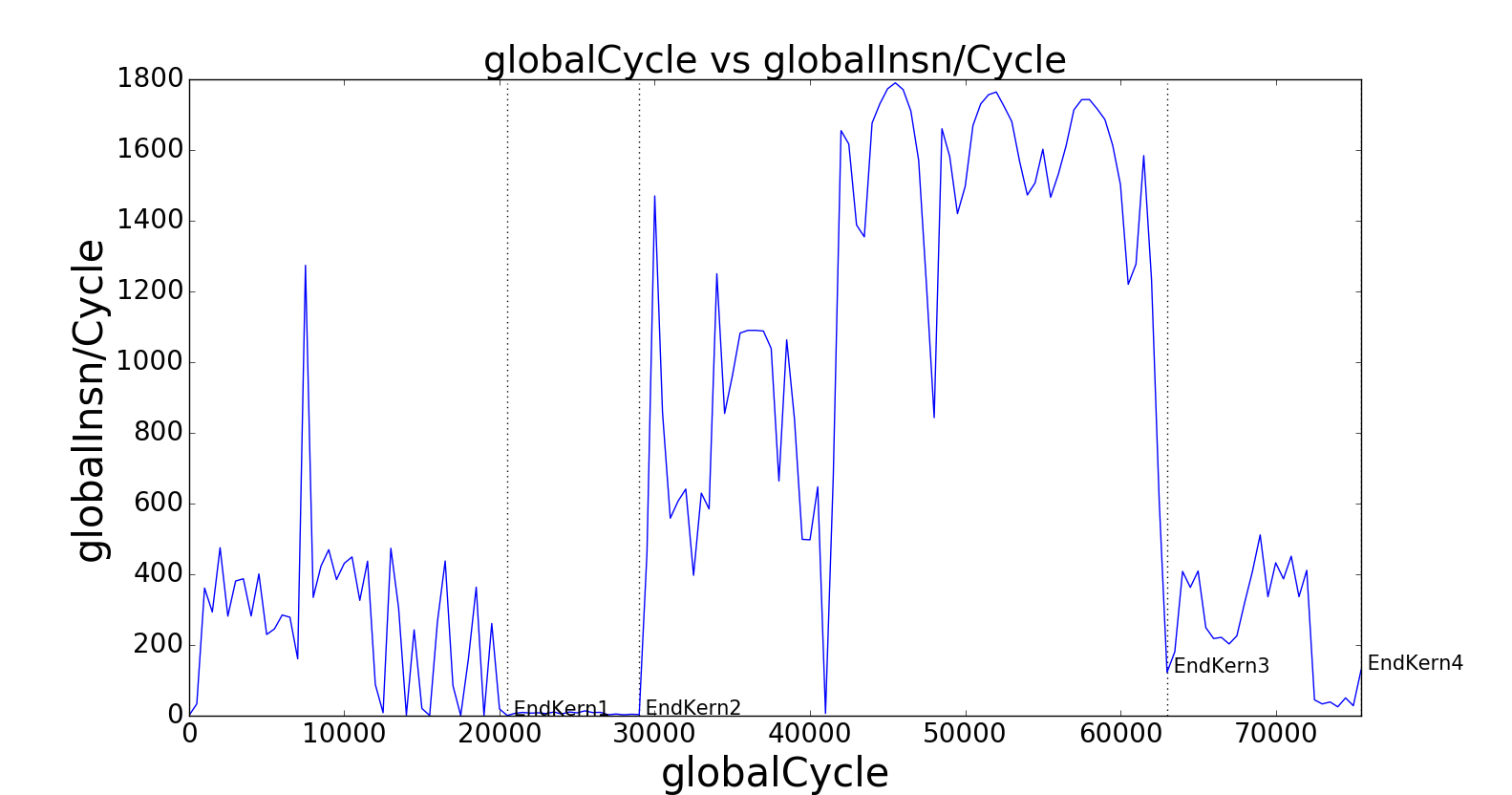}
  \caption{Forward Convolution (Winograd Nonfused) Global IPC Plot}
  \label{fig:gipc_fwd}
\end{figure}

\begin{figure}[tb!]
  \includegraphics[width=\linewidth]{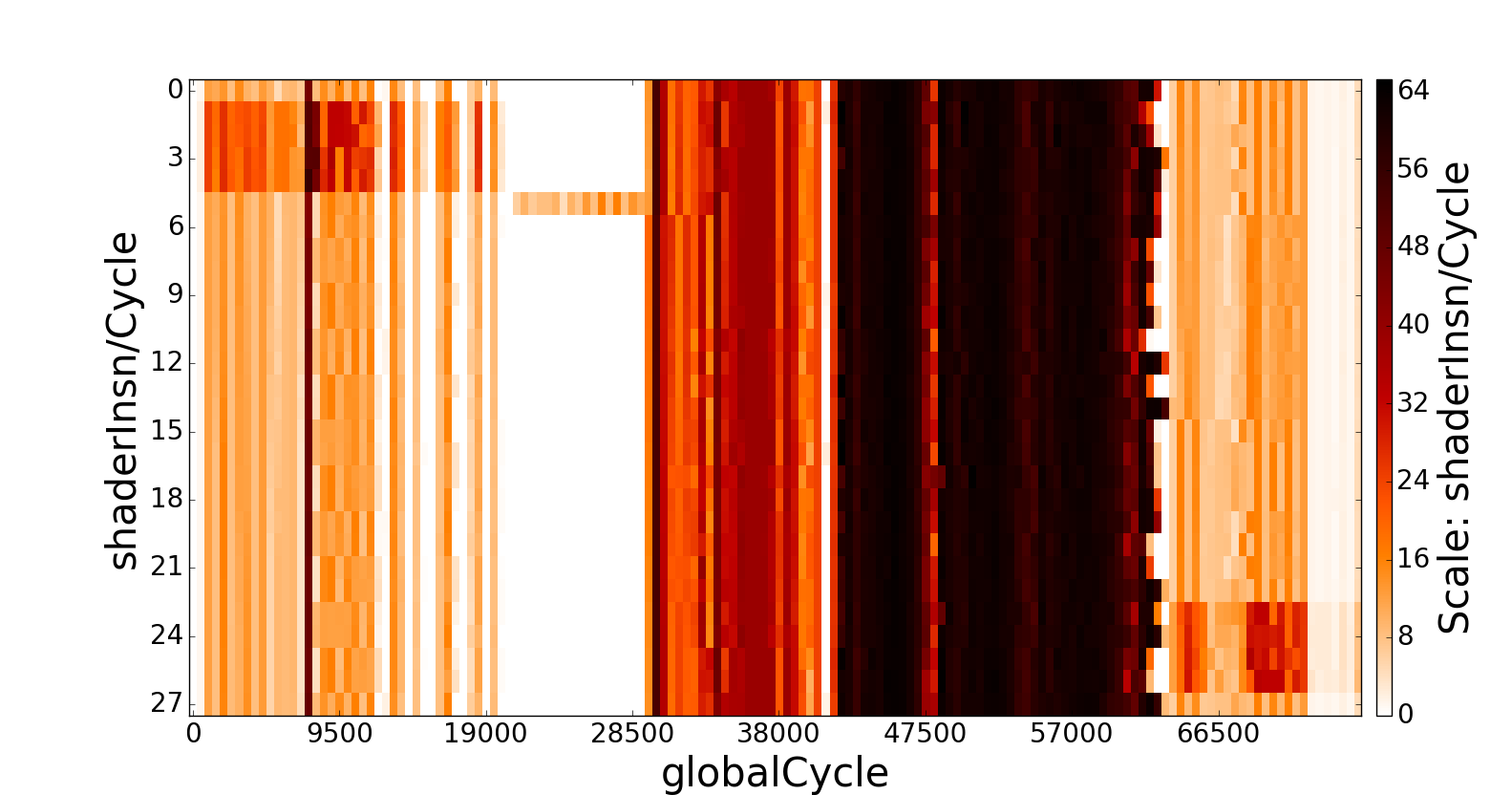}
  \caption{Forward Convolution (Winograd Nonfused) Shader IPC Plot}
  \label{fig:sipc_fwd}
\end{figure}

\begin{figure}[tb!]
  \includegraphics[width=\linewidth]{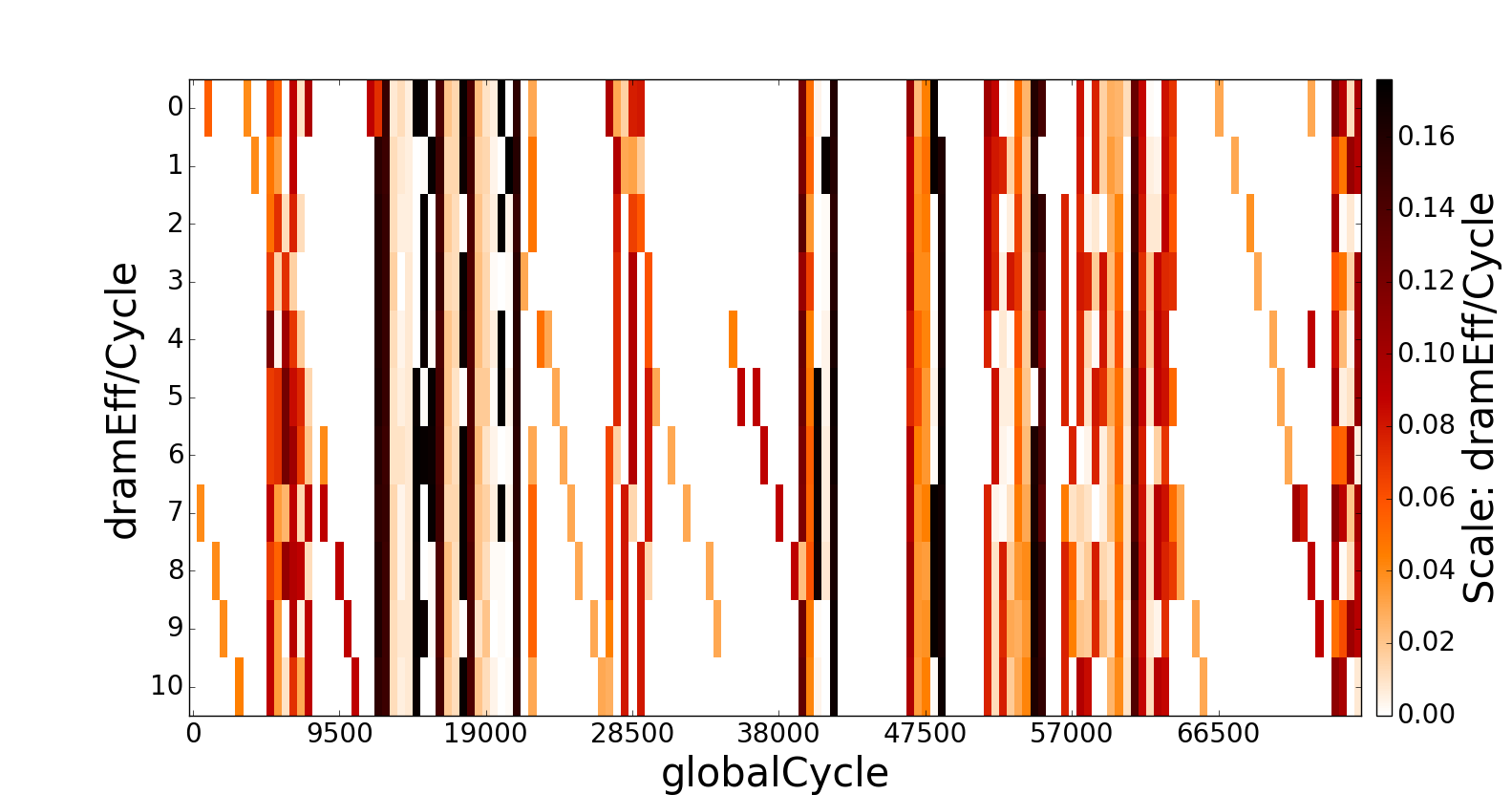}
  \caption{Forward Convolution (Winograd Nonfused) DRAM Efficency Plot}
  \label{fig:dram_fwd}
\end{figure}

\begin{figure}[tb!]
  \includegraphics[width=\linewidth]{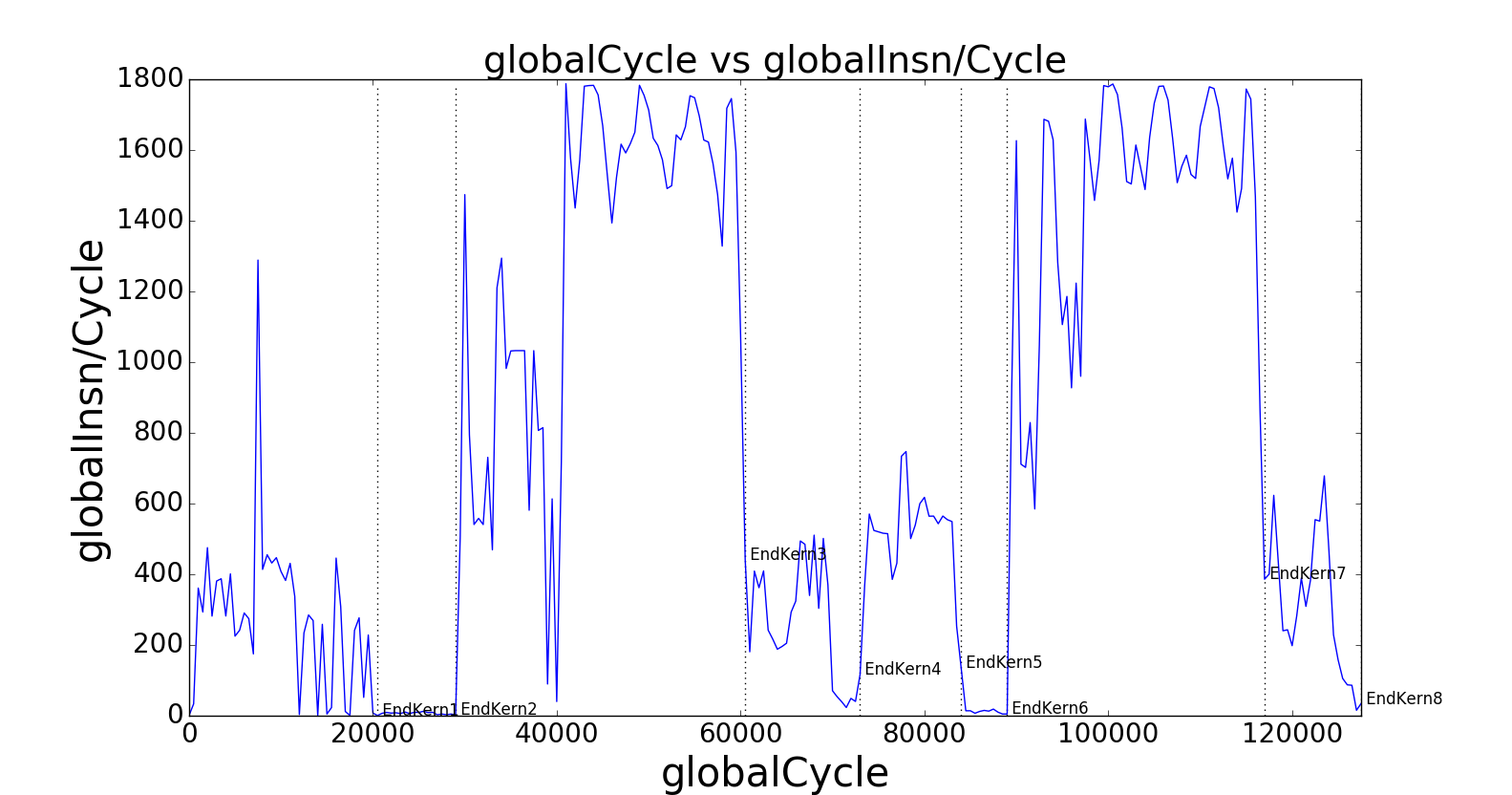}
  \caption{Backwards Data Convolution (Winograd Nonfused) Global IPC Plot}
  \label{fig:gipc_dgrad}
\end{figure}

\begin{figure}[tb!]
  \includegraphics[width=\linewidth]{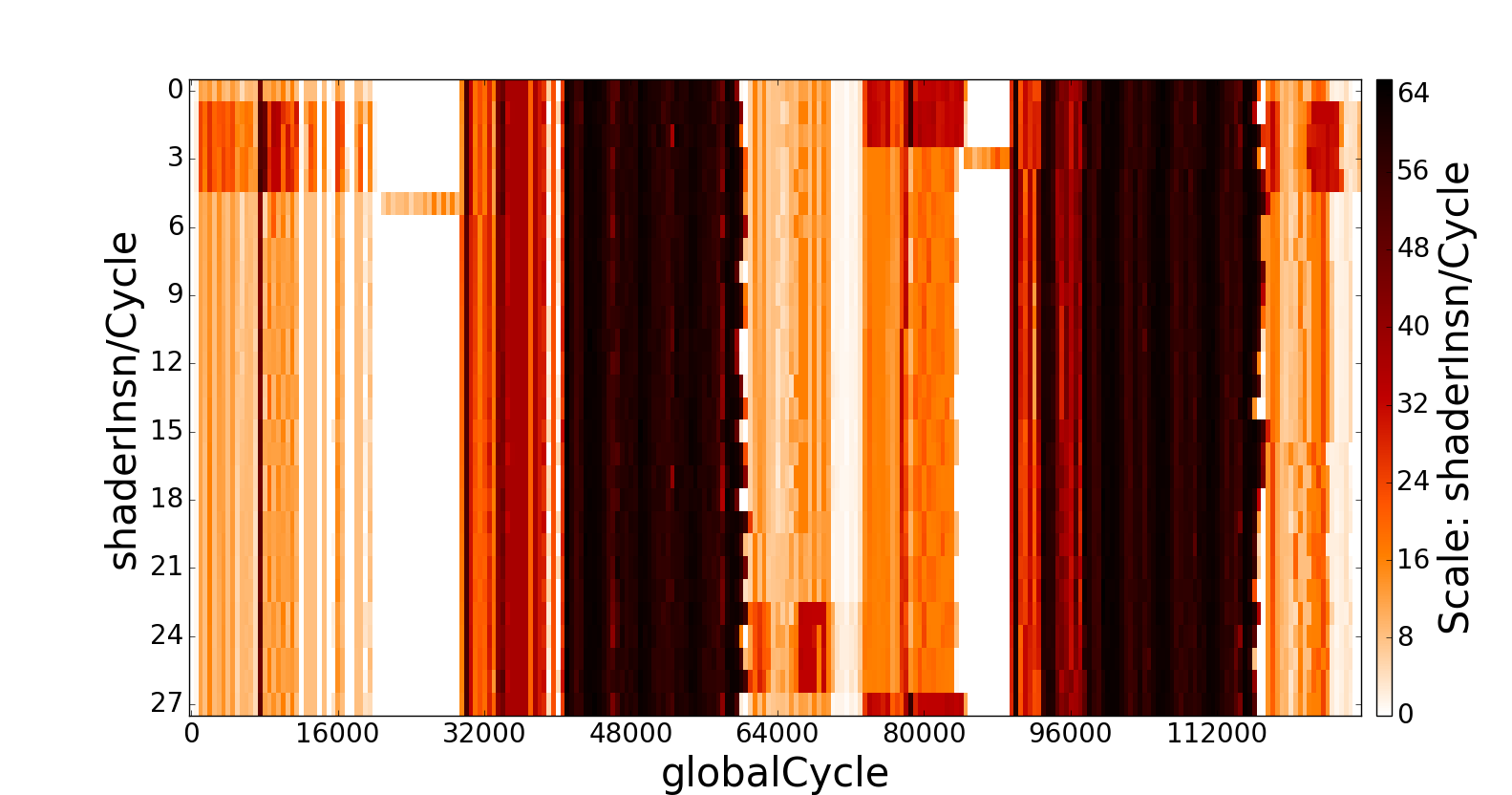}
  \caption{Backwards Data Convolution (Winograd Nonfused) Shader IPC Plot}
  \label{fig:sipc_dgrad}
\end{figure}

\begin{figure}[tb!]
  \includegraphics[width=\linewidth]{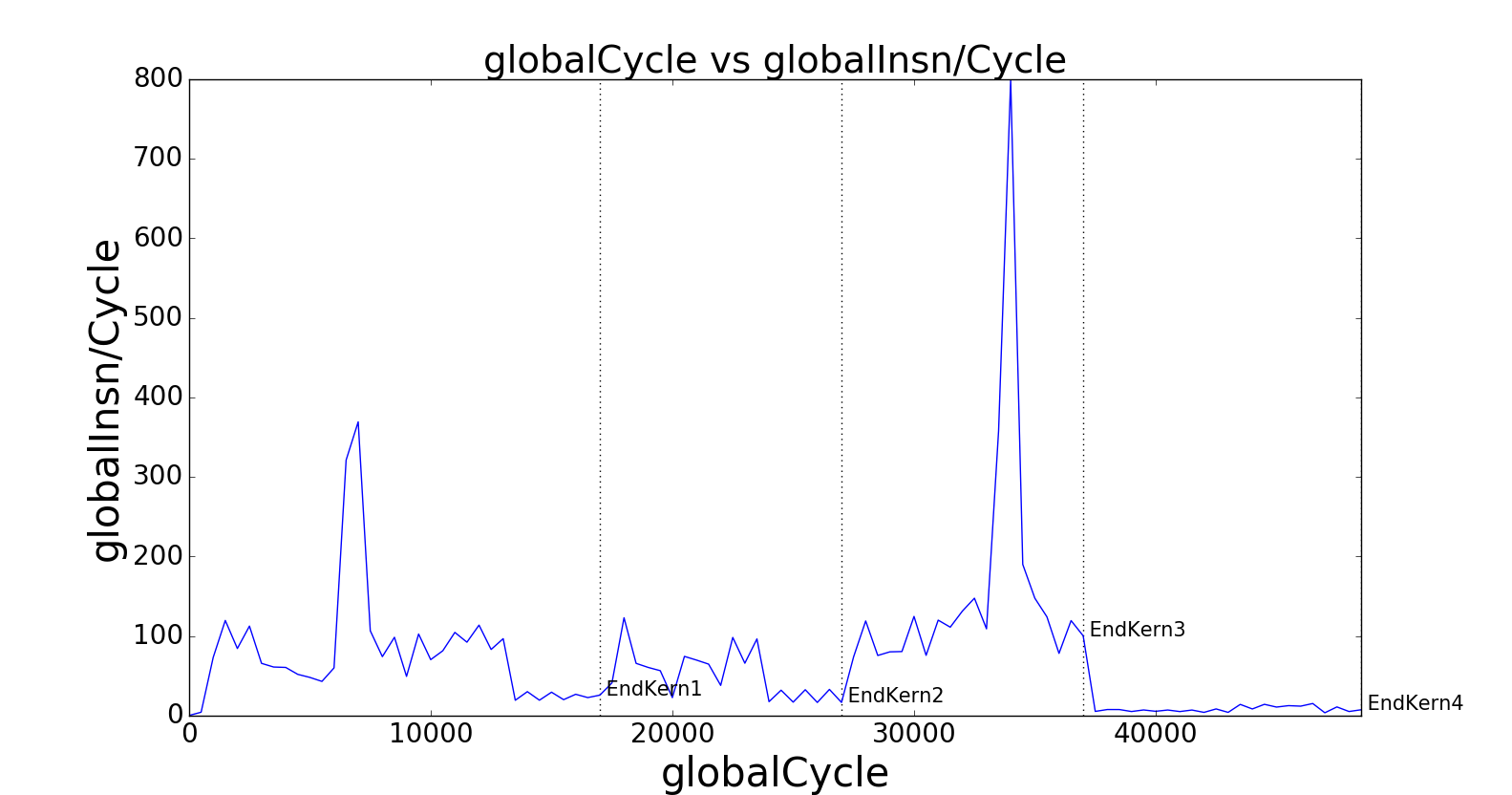}
  \caption{Backwards Filter Convolution (Winograd Nonfused) Global IPC Plot}
  \label{fig:gipc_wgrad}
\end{figure}

\begin{figure}[tb!]
  \includegraphics[width=\linewidth]{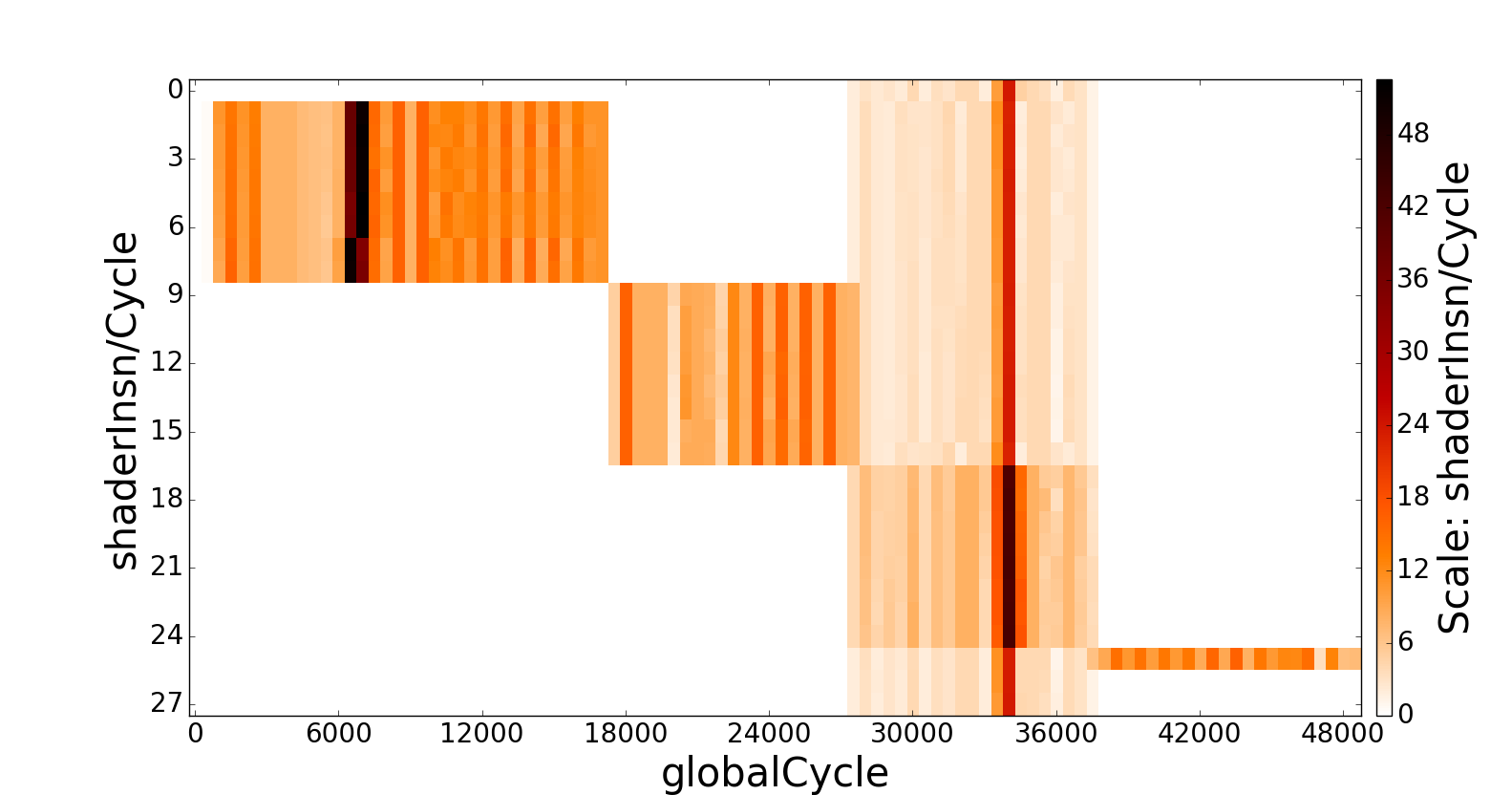}
  \caption{Backwards Filter Convolution (Winograd Nonfused) Shader IPC Plot}
  \label{fig:sipc_wgrad}
\end{figure}

\subsection{Global and Shader IPC}
\label{subsec:cases-ipc}

The Winograd Nonfused algorithm has the highest IPCs for all three types of convolution.  Furthermore, Figures~\ref{fig:gipc_fwd}, \ref{fig:sipc_fwd}, \ref{fig:gipc_dgrad}, and \ref{fig:sipc_dgrad} show that the forward convolution and backward data convolution implementations are balanced across all the shader cores and thus achieve high per shader IPCs. Although the backward filter convolution version of Winograd Nonfused, shown in Figures~\ref{fig:gipc_wgrad} and \ref{fig:sipc_wgrad}, still has the highest IPC, only some of the cores are being used due to load imbalance. However, for the active cores, it commits many instructions per cycle.  In general, the algorithms exhibit several clear phases.  For example, in Figure~\ref{fig:sipc_fwd}, only one core is actively committing instructions for a portion of the cycles and at other times, all cores are committing a lot of instructions quickly.  Thus, these results show that there are opportunities to save/reduce power by turning off cores during the phases they are not used.

\begin{figure}[tb!]
  \includegraphics[width=\linewidth]{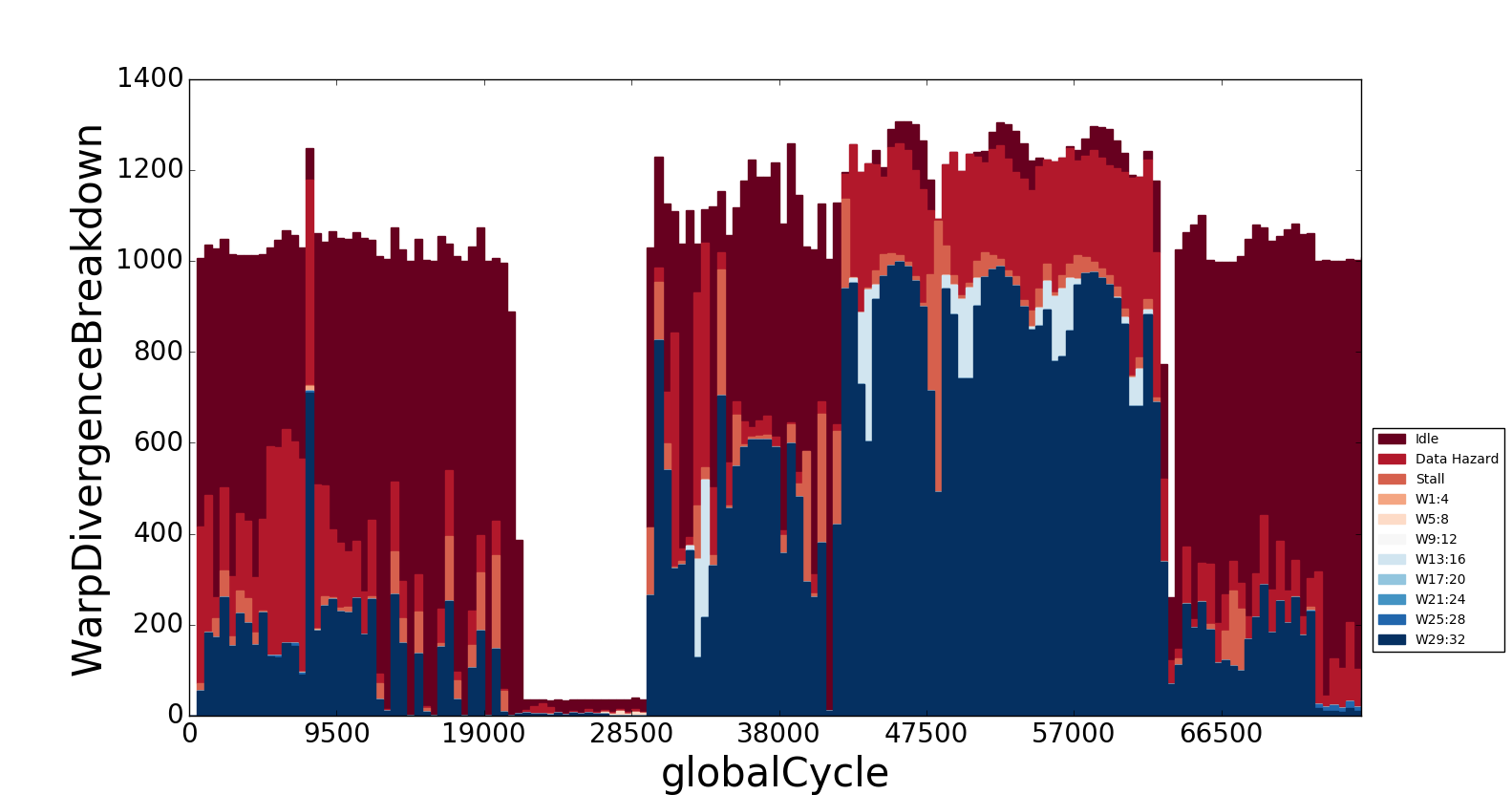}
  \caption{Forward Convolution (Winograd Nonfused) Warp Divergence Plot}
  \label{fig:warp_win}
\end{figure}

\begin{figure}[tb!]
  \includegraphics[width=\linewidth]{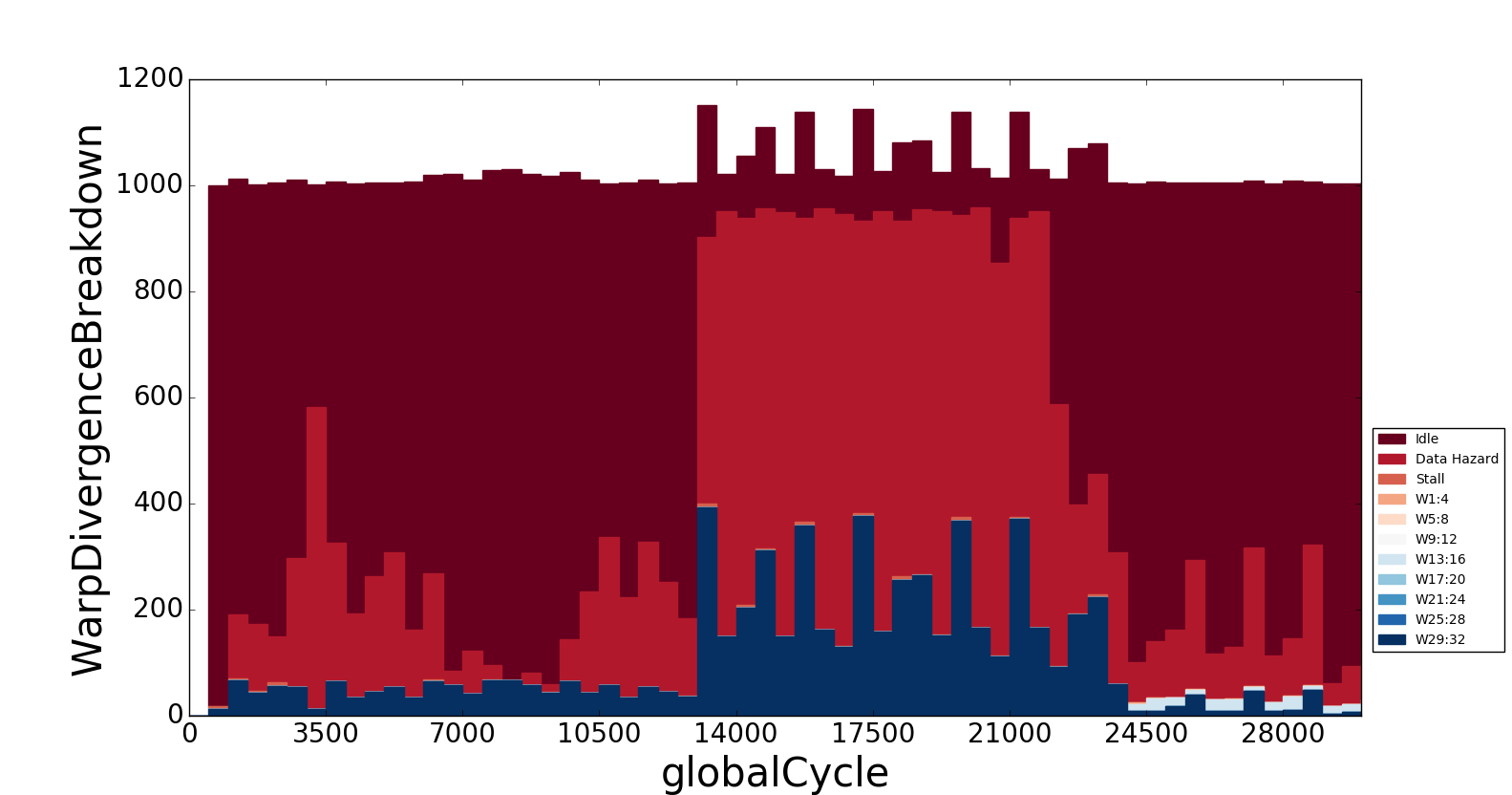}
  \caption{Forward Convolution (Implicit GEMM) Warp Divergence Plot}
  \label{fig:warp_imp}
\end{figure}

\begin{figure}[tb!]
  \includegraphics[width=\linewidth]{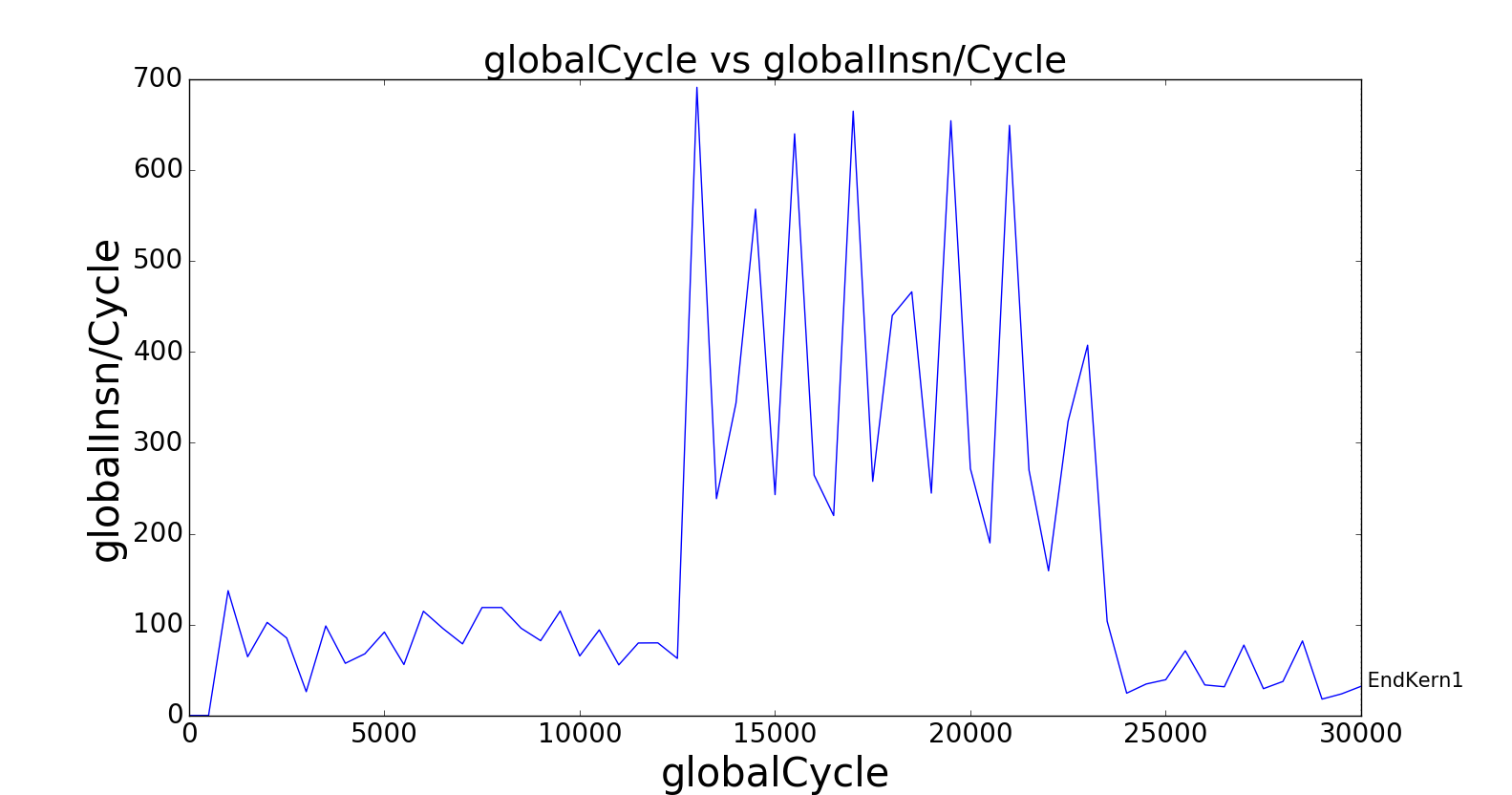}
  \caption{Forward Convolution (Implicit GEMM) Global IPC Plot}
  \label{fig:gipc_imp}
\end{figure}

\begin{figure}[tb!]
  \includegraphics[width=\linewidth]{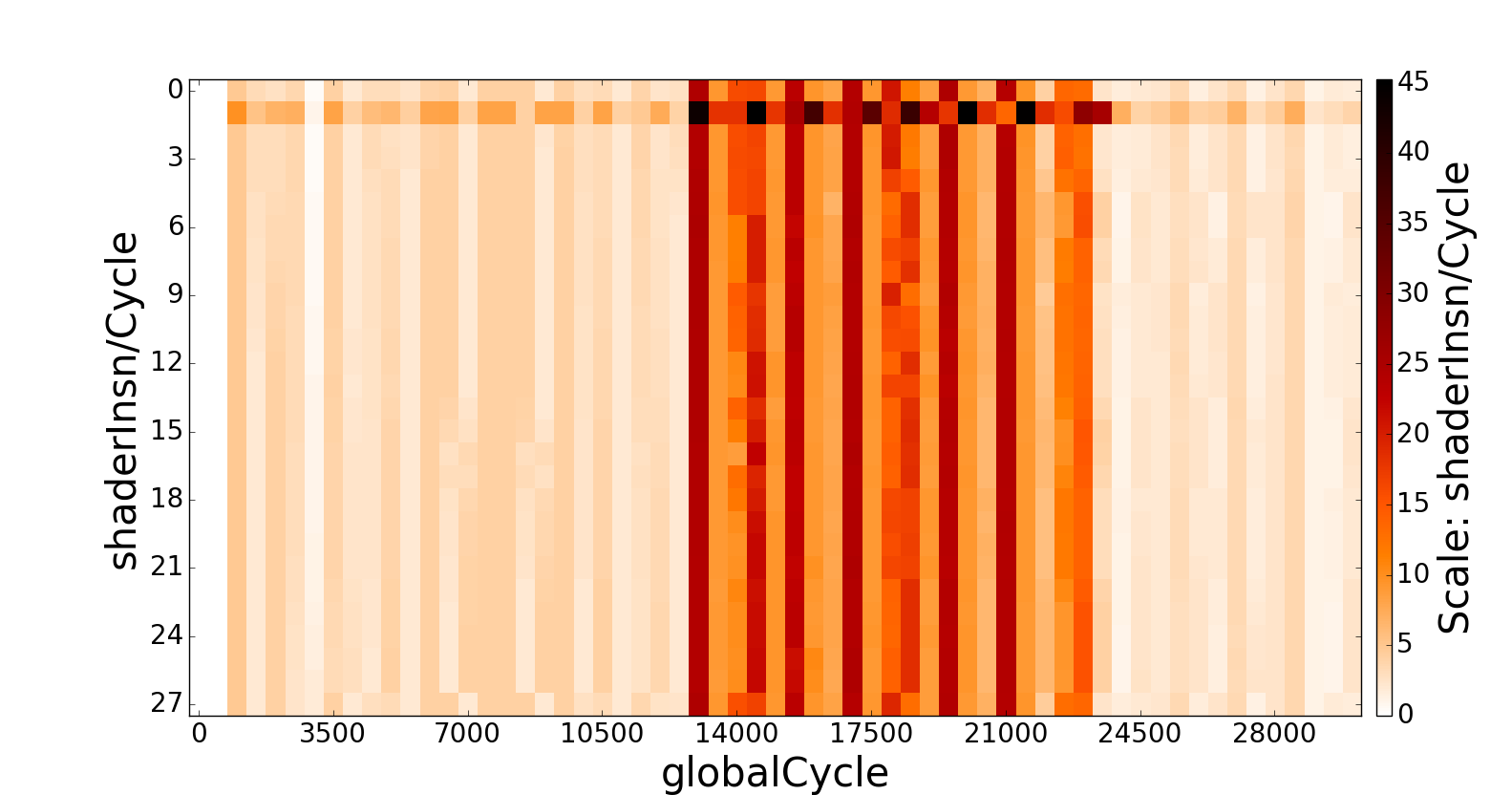}
  \caption{Forward Convolution (Implicit GEMM) Shader IPC Plot}
  \label{fig:sipc_imp}
\end{figure}

\subsection{Comparing DRAM Efficiency to IPC}
\label{subsec:cases-effic}

Figures~\ref{fig:sipc_fwd} and \ref{fig:dram_fwd} show that when Winograd Nonfused's IPC is highest, the memory efficiency is low, indicating that there are phases that the program is compute bound. For example, from around cycle 60000 to 70000, the execution is likely compute bound.  This demonstrates how GPGPU-Sim (and AerialVision) can be used to identify regions of interest in applications and how the GPGPU-Sim's statistics can enable detailed comparisions.

\subsection{Warp Divergence}
\label{subsec:cases-warpDiv}

In general, warp divergence is not an issue for any of the algorithms we tested -- likely because these algorithms are usually compute bound and have few branches. 
The forward convolution component of the Winograd Nonfused algorithm has the most significant warp divergence, as shown in Figure~\ref{fig:warp_win}, where up to two warps are executing at the same time.  However, this has a negligible impact on the IPC, since forward convolution with Winograd Nonfused is actually one of the fastest algorithms.  Nevertheless, this represents another piece of information that GPGPU-Sim provides and can be used to optimize other machine learning algorithms.

In Figure~\ref{fig:warp_imp}, we see that a majority of the warp breakdown is taken up by data hazards and idle warps. Comparing this to the IPC plots, Figures~\ref{fig:gipc_imp} and \ref{fig:sipc_imp}, we see that the low IPC despite the good load balance in the early part of the kernel can be attributed to this idle warp breakdown.  Thus, this represents another opportunity for optimization or power savings.

%% file: future.tex
\vspace{-1ex}
\section{Future Work}
\label{sec:future}


In addition to getting TensorFlow to run in GPGPU-Sim and adding complete FP16 support, the ability to capture the inputs to kernels requires further effort.  For example, our new GPGPU-Sim model can be enhanced to work with double pointer arguments.  Resolving these issues will allow us to extract specific kernels, run them individually on hardware, and study them using higher-level tools like NVProf.

%% file: conc.tex
\vspace{-1ex}
\section{Conclusion}
\label{sec:conc}

In this paper, we described the changes we made to the GPGPU-Sim simulator~\cite{aamodt2012gpgpu,bakhoda2009analyzing} to enable it to run PyTorch by running PTX kernels included in NVIDIA's cuDNN~\cite{chetlur2014cudnn} library. We use the resulting modified simulator, which we plan to make available publicly with this paper, to study deep learning workloads and analyze their behavior.  
With our changes to GPGPU-Sim's Functional simulation model, we find that GPGPU-Sim's Performance model running a cuDNN enabled implementation of LeNet for MNIST reports results within 30\% of real hardware. Using GPGPU-Sim's AerialVision performance analysis tool we observe that cuDNN API calls contain many varying phases and appear to include potentially inefficient microarchitecture behavior such as DRAM partition bank camping, at least when executed on GPGPU-Sim's current performance model. Since most deep neural networks deployed today are trained using GPUs via high-level frameworks such as TensorFlow~\cite{tf} and PyTorch~\cite{paszke2017automatic}, this work has the potential to enable significant microarchitectural research into current deep neural networks.